\documentclass[iop]{emulateapj}

\usepackage{natbib}
\usepackage{color}
\usepackage{graphicx}
\usepackage{epstopdf}

\newcommand{\Mjup}{\ensuremath{\mathrm{M}_{\mathrm{Jup}}\!}}
\newcommand{\Msun}{\ensuremath{\mathrm{M}_{\odot}\!}}

\newcommand{\Kfilt}{\ensuremath{\mathrm{K}}}
\newcommand{\Kpfilt}{\ensuremath{\mathrm{K}^{\prime}}}
\newcommand{\Ksfilt}{\ensuremath{\mathrm{K}_\mathrm{s}}}
\newcommand{\Kcfilt}{\ensuremath{\mathrm{K}_\mathrm{c}}}
\newcommand{\Hfilt}{\ensuremath{\mathrm{H}}}
\newcommand{\Jfilt}{\ensuremath{\mathrm{J}}}

\renewcommand{\deg}{\ensuremath{^{\circ}}}
\renewcommand{\arcsec}{\ensuremath{^{\prime\prime}}}

\newcommand{\unit}[1]{\ensuremath{\,\textrm{#1}}}

\slugcomment{Accepted for publication in ApJ}

\shorttitle{Friends of Hot Jupiters II: Direct Imaging}
\shortauthors{Ngo et al.}

\begin{document}

\title{Friends of Hot Jupiters II: No Correspondence Between Hot-Jupiter Spin-Orbit Misalignment and the Incidence of Directly Imaged Stellar Companions}

\author{Henry Ngo\altaffilmark{1}, Heather A. Knutson\altaffilmark{1}, Sasha Hinkley\altaffilmark{1,2}, Justin R. Crepp\altaffilmark{3}, Eric B. Bechter\altaffilmark{3}, Konstantin Batygin\altaffilmark{1}, Andrew W. Howard\altaffilmark{4}, John A. Johnson\altaffilmark{5}, Timothy D. Morton\altaffilmark{6}, and Philip S. Muirhead\altaffilmark{7}}
\email{hngo@caltech.edu}

\altaffiltext{1}{Division of Geological and Planetary Sciences, California Institute of Technology, Pasadena, CA, USA}
\altaffiltext{2}{Department of Physics and Astronomy, University of Exeter, Exeter, UK}
\altaffiltext{3}{Department of Physics, University of Notre Dame, Notre Dame, IN, USA}
\altaffiltext{4}{Institute for Astronomy, University of Hawaii at Manoa, Honolulu, HI, USA}
\altaffiltext{5}{Harvard-Smithsonian Center for Astrophysics, Cambridge, MA, USA}
\altaffiltext{6}{Cahill Center for Astronomy and Astrophysics, California Institute of Technology, Pasadena, CA, USA}
\altaffiltext{7}{Department of Astronomy, Boston University, Boston, MA, USA}

\begin{abstract}
Multi-star systems are common, yet little is known about a stellar companion's influence on the formation and evolution of planetary systems. For instance, stellar companions may have facilitated the inward migration of hot Jupiters towards to their present day positions.  Many observed short period gas giant planets also have orbits that are misaligned with respect to their star's spin axis, which has also been attributed to the presence of a massive outer companion on a non-coplanar orbit. We present the results of a multi-band direct imaging survey using Keck NIRC2 to measure the fraction of short period gas giant planets found in multi-star systems. Over three years, we completed a survey of 50 targets (``Friends of Hot Jupiters'') with 27 targets showing some signature of multi-body interaction (misaligned or eccentric orbits) and 23 targets in a control sample (well-aligned and circular orbits). We report the masses, projected separations, and confirmed common proper motion for the 19 stellar companions found around 17 stars. Correcting for survey incompleteness, we report companion fractions of $48\%\pm9\%$, $47\%\pm12\%$, and $51\%\pm13\%$ in our total, misaligned/eccentric, and control samples, respectively. This total stellar companion fraction is $2.8\,\sigma$ larger than the fraction of field stars with companions approximately $50-2000\unit{AU}$. We observe no correlation between misaligned/eccentric hot Jupiter systems and the incidence of stellar companions. Combining this result with our previous radial velocity survey, we determine that $72\% \pm 16\%$ of hot Jupiters are part of multi-planet and/or multi-star systems.
\end{abstract}


\section{Introduction}
Surveys of Solar-like stars within 25\unit{pc} indicate that approximately 44\% are found in multiple star systems~\citep{Raghavan2010}. At the same time, recent surveys have sought to quantify planet occurrence rates around solar-type FGK stars~\citep[e.g.][]{Howard2012b,Fressin2013}. However, the effects of additional stellar companions on the formation and subsequent evolution of planetary systems are not well understood. A stellar companion might disrupt planet formation by stirring up the disk~\citep[e.g.][]{Mayer2005}, truncating the disk~\citep[e.g.][]{Pichardo2005,Kraus2012}, or ejecting planets~\citep[e.g][]{Kaib2013,Zuckerman2014}. Numerical simulations often fail to produce planets in binary star systems~\citep[e.g.][]{Pichardo2005,Mayer2005,Thebault2006,Fragner2011}, suggesting that a stellar companion can indeed hinder planet formation. On the other hand, analytic calculations predict that stellar companions would have little effect on planetesimal growth~\citep[e.g.][]{Batygin2011,Rafikov2013a,Rafikov2013b} and current surveys have found a number of planets in binary star systems~\citep[e.g.][]{Eggenberger2007,Raghavan2010,Kaib2011,Orosz2012a,Orosz2012b}. In addition, a stellar companion might cause planets to migrate via three-body interactions, such as the Kozai-Lidov mechanism or via other secular interactions, resulting in very small orbital distances~\citep[e.g.][]{Malmberg2007a,Malmberg2007b,Fabrycky2007,Morton2011a,Naoz2012,Naoz2013,Teyssandier2013,Petrovich2015,Storch2014}. However, the Kozai-Lidov mechanism is suppressed in multi-planet systems because planet-planet interactions tend to prohibit the libration of the argument of perihelion characteristic of the Kozai-Lidov resonance~\citep{Wu2003,Batygin2011}. Finally, stellar companions can also bias our estimates of the properties of transiting planet systems by diluting the measured transit depth, resulting in an underestimate of the planet's radius and a corresponding overestimate of its density.

In this study we focus on a class of short-period gas giant planets known as ``hot Jupiters.'' These planets could not have formed at their current locations, but must have migrated in from beyond the ice lines of their natal disks~\citep[e.g.][]{Lin1996}. However, the mechanism(s) responsible for hot Jupiter migration remain controversial. Current migration models include disk interactions~\citep[e.g.][]{Goldreich1980,Tanaka2002,Lin1986} and gravitational interaction with a third body, such as another planet~\citep[e.g.][]{Chatterjee2008,Nagasawa2008,Wu2011,Beauge2012,Lithwick2014} or a stellar companion. In general, isolated simple disk migration models produce hot Jupiters on circular orbits that are well aligned with the primary star's spin axis, while migration due to a third body leads to hot Jupiters that are often eccentric and/or misaligned with the primary's star spin axis. 

Surveys from the past few years~\citep[e.g.][]{Winn2010b,Albrecht2012} indicate misaligned hot Jupiters are common--18 out of the 53 hot Jupiters surveyed to date have obliquities that are inconsistent with zero at the three sigma level or higher. As a result, it has been argued that a significant fraction of hot Jupiters may have migrated via three-body interactions such as the the Kozai-Lidov effect, which naturally results in large orbital inclinations~\citep[e.g.][]{Morton2011a,Li2014}. If stellar tides can bring misaligned hot Jupiters back into alignment with the star's spin axis~\citep{Dawson2014a}, this fraction may be even higher than the current rate suggests. Conversely, \citet{Dawson2015} argue that the lack of high eccentricity Jupiters at intermediate periods in the overall {\it Kepler} sample places a strict upper limit on the fraction of hot Jupiters that might have migrated via three-body processes. Misaligned hot Jupiters may also result from migration in a tilted disk, which could be caused by torque from a distant stellar companion~\citep{Batygin2012}. Moreover, significant star-disk misalignments may naturally arise from the physical evolution of the star and the disk in a perturbed system~\citep{Batygin2013,Spalding2014}. This suggests that a hot Jupiter's obliquity, which can be measured with the Rossiter-McLaughlin effect~\citep{Winn2005} or via Doppler tomography~\citep[e.g.][]{CollierCameron2010b,Brown2012}, might provide a clue to whether or not a third body has influenced the planetary system. 

Alternatively, planet-planet scattering could produce misaligned hot Jupiters without requiring the presence of a stellar companion. \citet{Dawson2013} also find evidence that high-eccentricity proto-hot Jupiters are more common around metal-rich stars, which presumably are more likely to have formed multiple gas giant planets. Other studies have also suggested that protoplanetary disks in isolation might in fact be tilted by the chaotic nature of star formation~\citep{Bate2010}, the primary star's magnetic torques~\citep{Lai2011}, and stellar surface modulation by internal gravity waves~\citep{Rogers2012,Rogers2013b}.

If a significant fraction of hot Jupiters migrate inward and acquire spin-orbit misalignments via three-body interactions, then this necessarily requires the presence of a massive outer planetary or stellar companion in these systems.  However, there have not been any studies published to date that have provided a well-constrained estimate of the frequency of bound stellar companions in hot Jupiter systems. A few stellar companions to transiting planet host stars were discovered serendipitously as part of studies intended to better characterize the transiting planet and its host star~\citep[e.g.][]{CollierCameron2007,Crossfield2012,Sing2013}. Some other works, such as \citet{Narita2010b,Narita2012} report directly imaged stellar companions from adaptive optics (AO) follow-up of known planets but only for one or two transiting gas giant planetary systems. The first systematic surveys for stellar companions to transiting planet systems used the ``Lucky imaging'' method. These studies focused exclusively on transiting hot Jupiter systems and their sample sizes were small: 14 in \citet{Daemgen2009}, 16 in \citet{Faedi2013} and 21 in \citet{Bergfors2013}. In addition, many of these surveys observed overlapping target lists.

More recently, there have been a series of studies focusing on the sample of {\it Kepler} transiting planet candidate host stars. The two surveys by ~\citet{LilloBox2012,LilloBox2014}, covering 174 {\it Kepler} planet host candidates, is the largest ``Lucky Imaging'' search to date. Current state-of-the-art direct imaging surveys use adaptive optics to achieve diffraction-limited imaging to allow for better detection and survey efficiency. \citet{Adams2012}, \citet{Adams2013}, \citet{Dressing2014}, and \citet{Wang2014} obtained infrared AO images of 90, 12, 87, and 56 {\it Kepler} planet candidate hosts, respectively. \cite{Adams2013} also searched around 15 transiting planet hosts. \citet{Law2014} recently published the first part of an optical campaign to search for companions around all {\it Kepler} planet candidate hosts using the Robo-AO instrument, with an initial sample size of 715 stars. \citet{Gilliland2015} searched for companions around 23 {\it Kepler} planet candidate hosts using optical images from the Hubble Space Telescope WFC3 instrument. Finally, \citet{Horch2014} used differential speckle imaging in two optical bandpasses to search for companions around 623 {\it Kepler} planet candidate hosts.

Unlike previous surveys of hot Jupiters detected by ground-based transit surveys, these imaging surveys were intended to confirm the planetary nature of the transits detected by {\it Kepler} and to correct the transit light curves for any dilution due to nearby stars, therefore ensuring accurate planetary radius estimates. Because the typical proper motions of the {\it Kepler} stars are quite small, these studies report relative brightness and projected separation for companions but do not attempt to determine whether or not they are bound companions or background objects.  The planetary systems in these surveys have a size distribution that reflects that of the {\it Kepler} survey as a whole, with the majority of systems consisting of sub-Neptune-sized transiting planets.

In this work, we present a diffraction-limited direct imaging survey of close-in transiting gas giant planets orbiting bright, nearby stars, as part of the ``Friends of Hot Jupiters'' campaign. These systems are among the most favorable targets for the Rossiter-McLaughlin technique, and the majority of our targets have published measurements of their spin-orbit alignment.  By focusing on this sample, we can directly test current hot Jupiter migration models and investigate the origin of their observed spin-orbit misalignments by searching for massive, distant companions in these systems. Our survey uses multiple bandpasses and repeated observations spanning a several year baseline in order to determine whether any directly imaged companions are physically bound. We also use these same data to estimate companion masses and projected physical separations, which are required in order to evaluate the likelihood of specific dynamical evolution scenarios for these systems.

The Friends of Hot Jupiters survey uses multiple companion detection modes to search for planetary and stellar companions around exoplanetary systems. Our sample consists of fifty-one exoplanetary systems that are known to host a transiting gas giant planet with masses between $0.06-11\,\Mjup$ and periods between $0.7-11$ days. We divide this sample into two sub-samples, consisting of planets that are on misaligned and/or eccentric orbits and a control sample of planets on apparently circular, well-aligned orbits~\citep[see][for a full description of the sample selection for this survey]{Knutson2014}. We consider targets to be ``misaligned'' if they host planets with an eccentricity or spin-orbit alignment more than 3 standard deviations away from zero. In \citet{Knutson2014}, we presented our search for long-term radial velocity (RV) accelerations due to distant massive planetary or stellar companions in these systems. We found a total companion occurrence rate of $51\%\pm10\%$ for companions with masses between $1-13\unit{\Mjup}$ and orbital semimajor axes between $1-20\unit{AU}$, with no evidence for a higher frequency of radial velocity companions in systems with eccentric and/or misaligned gas giant planets. In a future paper we will present the results of a complementary search for close-in stellar companions using high resolution K band spectroscopy, which is primarily sensitive to K and M stars within 0.5\arcsec of the primary.

In this paper, we present the results of our diffraction-limited direct imaging search. In Section~\ref{sec:obs}, we describe our observations. In Section~\ref{sec:analysis}, we summarize the point spread function (PSF) fitting method used to calculate the brightness ratio and positions of the candidate stellar companions, as well as upper limits for companions in systems with non-detections. We then determine whether or not the candidate companions share common proper motion with the primary, and estimate their projected physical separations and masses. In Section~\ref{sec:companions}, we discuss each system individually. In Section~\ref{sec:compfrac}, we compare our estimated frequency for stellar companions to the results from previous surveys of planet-hosting and field star samples. Finally, in Section~\ref{sec:summary}, we summarize our findings and discuss the implications of our measured companion fraction for the formation of hot Jupiter systems.

\section{Observations}
\label{sec:obs}
During the AO phase of our survey, we collected data for 50 out of 51 FHJ systems with the NIRC2 instrument (instrument PI: Keith Matthews) on Keck II using K band natural guide star adaptive optics imaging. We were not able to image one target, WASP-19, because its declination, $-45.7\deg$, was too far south to observe with Keck AO. Two of our target systems, HAT-P-8 and WASP-12, turned out to be triple systems, which we previously reported in \citet{Bechter2014}. We obtained observations between February 2012 and October 2014 and our observations are summarized in Table~\ref{tab:obs}. We used the full array (1024x1024 pixel field of view) on the narrow camera setting (10$\,$mas$\,$pixel$^{-1}$) to maximize our spatial resolution. However, for several bright targets (as noted in Table~\ref{tab:obs}), we used a subarray to reduce integration times and avoid saturation. We used a three-point dither pattern to reduce the effects of the NIRC2 array's noisier lower left quadrant and instrumental noise levels while also preserving our sensitivity to companions with higher spatial separations. We aimed for a total of two minutes of on-target integration time per system in position angle mode, where the orientation of the image is kept constant on the detector as the telescope tracks. This technique allows us to detect companions with $\Delta K$ of approximately $8$ at separations of approximately $1\arcsec$. For targets where a potential companion object is seen, we repeat the observations in at least one other filter, such as J or H, in order to obtain color information. We also follow up on targets with detected companions approximately one or more years later to obtain K band astrometric measurements necessary to confirm that the companion is gravitationally bound via a common proper motion analysis. We elect to use K band rather than J or H band for our astrometry because the AO correction is superior at longer wavelengths.

We calibrate our images using dome flat fields and dark frames. We also find and remove image artifacts. We flag flat field pixels that are less than 0.1 times the median as dead pixels and dark frame pixels that are more than $10\sigma$ from the median as hot pixels. For each frame, we identify the remaining bad pixels as those with counts that are $8\sigma$ outliers compared to the counts in pixels in the surrounding 5x5 box. We replace all the flagged pixels' value with the median of the 5x5 box centered on the flagged pixel. We use these calibrated individual frames in all of our photometric and astrometric analyses. We limit our integrations to stay just below the nonlinear regime for the NIRC2 detector, and use Poisson statistics to determine the uncertainty in our counts. We also create a single, reduced image by aligning the individual frames so that the target star is in the same position and then combine using a median stack. We use the stacked image for our sensitivity calculations.

\section{Analysis of Companion Properties}
\label{sec:analysis}
\subsection{Detections}
\label{sec:psffit}
\begin{figure*}
\epsscale{1.18}
\plotone{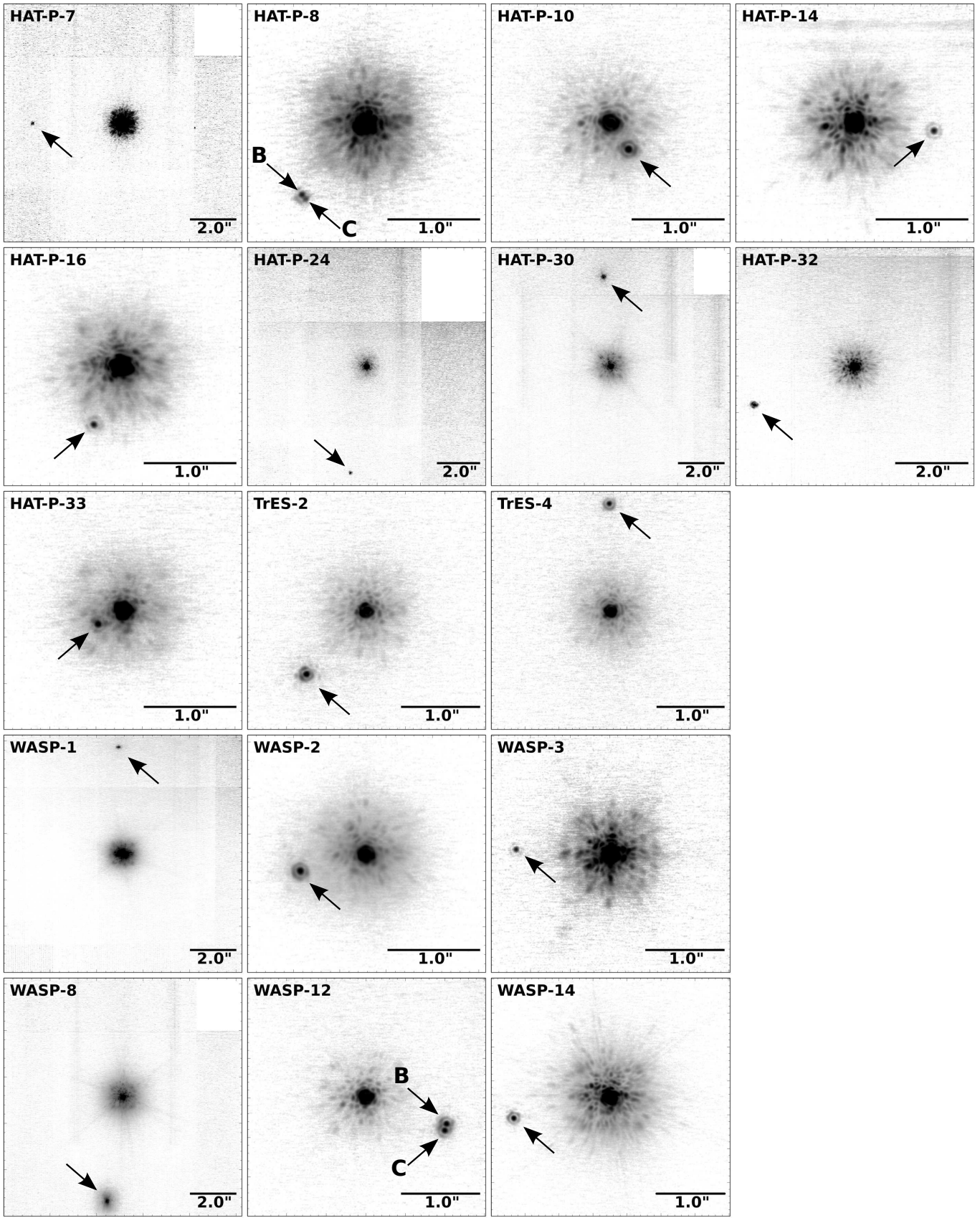}
\caption{Median-stacked K-band images showing the Friends of Hot Jupiters survey targets with detected and confirmed stellar companions (marked with arrows). The first epoch observation (see Table~\ref{tab:obs}) for each target was used.
\label{fig:comps}}
\end{figure*}

We find 15 binary systems and 2 triple systems, for a total of 17 multi-star systems, out of the 50 systems with AO observations. We show one median-stacked K-band image for each of these detections in Figure~\ref{fig:comps}. Table~\ref{tab:stellar_params} summarizes the stellar parameters for all Friends of Hot Jupiters survey targets and the number of companions found around each star.

To measure the flux ratio and on-sky separations for each system, we fit a multiple-source point spread function (PSF) to each calibrated frame. Following~\citet{Bechter2014}, we choose to model the PSF with as a Moffat function with a Gaussian component,
\begin{equation}
\footnotesize
I(x,y) = \sum_{i=1}^{N_{*}} \left( \alpha_i \left[1 + \left(\frac{r_i}{r_s}\right)^2 \right]^{-\beta}+\gamma_i \exp\left[-\frac{r_i^2}{w^2}\right]\right) + b,
\label{eq:psf}
\end{equation}
where $N_{*}$ is the number of stars in the image (either 2 or 3); $r_i=\sqrt{x_i^2+y_i^2}$ is the distance from the $i$-th star; $x_i,y_i,\alpha_i, \gamma_i$ are parameters that vary with each star and determine the position of the star and the amplitude of the PSF; $\beta$ takes a single value for all stars and sets the exponent of the Moffat contribution; $r_s$ and $w$ each take a single value for all stars and determine the width of the Moffat and Gaussian portions of the PSF, respectively; and $b$ is the background sky level. Hence, the total number of free parameters is $4N_{*} + 4$. We also only fit a circular aperture of radius 10 pixels around each star. From experimenting with different aperture sizes, we find this radius covered most of the star's flux (the full width half maximum, FWHM, is about 5 pixels) while remaining small enough to avoid counting any remaining bad pixels in the background. We also explored some alternative PSF fitting schemes using a smaller sample from our survey. We tried a Moffat function combined with an elliptical Gaussian model, a purely Gaussian model, and a sinc$^2$ model. From examining the Bayesian Information Criteria, we find that the best model is the Moffat function with a radially symmetric (circular) Gaussian component.

We find the best fit parameters using a maximum likelihood estimation routine. These best fit parameters determine an analytic form for our PSF model. We compute the flux ratio by integrating the best-fit PSF model over the same circular aperture used in the PSF fitting for each star. We use the difference in the stellar position parameters to compute the horizontal and vertical separation as projected onto the NIRC2 array. We then adjust these separations to account for the well characterized distortion and rotation of the NIRC2 array using the astrometric corrections presented in~\citet{Yelda2010}. 

We compute the flux ratio and corrected one-dimensional separations for each frame. Our best estimate of these measurements for each observation is the median value from all of the frames. We calculate the standard error on the mean and use that as our measurement error. Using the best estimate for the one-dimensional separations and the corrected NIRC2 plate scale~\citep{Yelda2010}, we then compute the projected on-sky separation $\rho$ and position angle PA for each detected companion. The NIRC2 astrometric corrections include uncertainties on the distortion, plate scale, and the orientation of the NIRC2 array. Therefore, we include these uncertainties in our total error budget for our measured $\rho$ and PA. 

We complete the above analysis to determine the photometry for all detected companions in all bandpasses (\Jfilt, \Hfilt, \Kpfilt, \Ksfilt) and find the best fitting flux ratio between primary and companion stars in each band, reported as a difference in magnitudes in Table~\ref{tab:comp_phot}. We then compute the apparent magnitudes of the companion stars in all bands as well as the colors of the companion stars. In order to do this, we require the apparent magnitudes of the primary star, which we obtained from the 2MASS catalog~\citep{Skrutskie2006}. Table~\ref{tab:phot_colors} shows the full multiband photometry of our detected companion stars. For our astrometric analysis, we only use K-band data and report the best fit separations and position angles in Tables~\ref{tab:comp_astr}.

\subsection{Common proper motion confirmation}
Our next step is to determine whether or not our candidate companions share common proper motion with the primary star, indicating that they are bound companions rather than background sources in the same line of sight. If the detected companion is actually a very distant background star, it will remain effectively stationary while the closer primary star moves across the sky as dictated by its parallax and proper motion. We would therefore expect that a background object would display a time-varying separation and position angle relative to the primary star, while a bound companion will maintain a constant separation and position angle.

\begin{figure*}
\epsscale{1.18}
\plotone{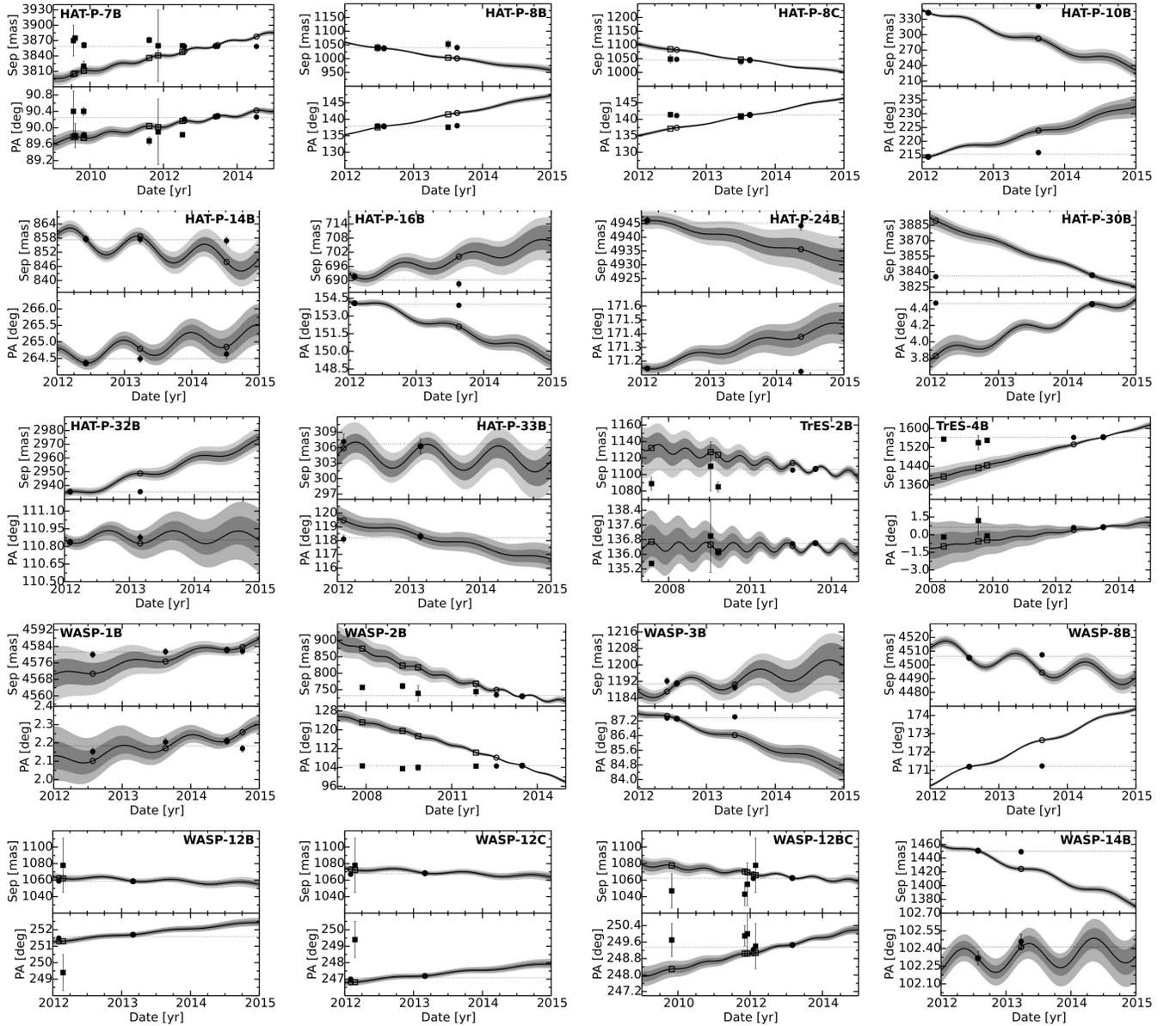}
\caption{For each object, the two panels show the separation (top) and position angle (bottom) of a detected companion star relative to the primary target star. The lines show the track of a background object, as computed from our observation with the smallest uncertainty. The dark- and light-gray shaded regions indicate the 68\% and 95\% confidence region for this track. Filled symbols indicate measured positions of companions while open symbols indicate the position the detected source would have if it were a background source. Measurements from this work are plotted as circles and shown in Table~\ref{tab:comp_astr}, while measurements from previous studies are plotted as squares and shown in Table~\ref{tab:prev_studies}. Detected companions can be ruled out as background objects if either their separation or position angle are inconsistent with the background track. The two triple systems, HAT-P-8 and WASP-12, have each of their companion candidates plotted separately. Because the individual analysis for WASP-12B and WASP-12C cannot conclusively confirm or rule out common proper motion, we also plot the astrometric measurements for the center of mass for the combined light of both companions as WASP-12BC. All objects in this figure are determined to be common proper motion companions to their primary star.
\label{fig:astr_plot}}
\end{figure*}

\begin{figure}
\epsscale{0.9}
\plotone{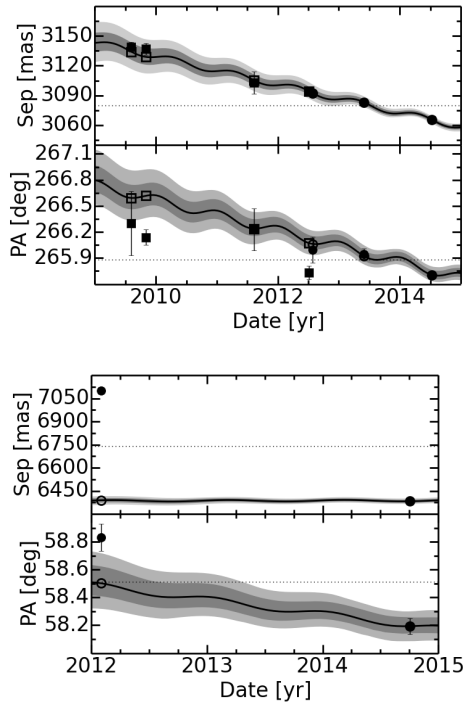}
\caption{These plots are the same as Figure~\ref{fig:astr_plot}, except this Figure shows the two candidate objects that were determined to be background objects. The measurements are reported in Table~\ref{tab:astr_bg}. The top plot shows the astrometric measurements for the second candidate companion to HAT-P-7, with circles showing our data and squares showing the four measurements for the ``western companion'' reported in \citet{Narita2012}. The bottom plot shows the astrometric measurements for the candidate companion to HAT-P-15. The data rules out a bound companion conclusion for both of these candidates.
\label{fig:astr_plot_bg}}
\end{figure}

We calculate the ``background track'' (i.e. the evolution of the companion's separation and position angle as a function of time if it was a background object) as shown in Figures~\ref{fig:astr_plot} and \ref{fig:astr_plot_bg}. We compute the primary star's parallactic motion using the celestial coordinates of the primary star from the SIMBAD database and the Earth ephemerides from the JPL Horizons service. The primary star's proper motion and uncertainties are also taken from the SIMBAD database. When determining the background tracks, we account for uncertainties in the primary star's celestial coordinates, proper motion, and parallax in addition to our measurement uncertainties in separation and position angle. We run a Monte Carlo routine to calculate the uncertainty in the background tracks. The 68\% and 95\% confidence regions for the separation and position angle evolution are shown as shaded regions in Figures~\ref{fig:astr_plot} and \ref{fig:astr_plot_bg}. We use the measurement with the smallest uncertainty in separation and position angle as the starting point for our track.

After creating these background tracks, we next overplot the measured companion separation and position angle at each epoch from Tables~\ref{tab:comp_astr}, ~\ref{tab:prev_studies}, and ~\ref{tab:astr_bg}. Several of our candidate companions were detected in previous imaging surveys; when available, we also show the separations and position angles from these earlier studies. We provide a complete list of these previously published detections in Tables~\ref{tab:prev_studies} and \ref{tab:astr_bg}, and discuss individual systems in more detail in section~\ref{sec:prev_studies}. In Figures~\ref{fig:astr_plot} and \ref{fig:astr_plot_bg}, measurements from this study are plotted as circles while measurements from other studies are plotted as squares. 

Based on these measurements, we conclude that all but two of our detected candidate companions must be bound to the host star.  The exceptions are the second candidate companion seen around HAT-P-7 and the candidate companion seen around HAT-P-15. Our followup observations determined these candidate companions to be background objects. Due to the small projected physical and angular separations of our companion candidates, our result that the majority of our candidates are physically bound companions is consistent with other direct imaging surveys that confirm association via multi-epoch detections~\citep[e.g.][]{Eggenberger2007,Bowler2015} or via galactic crowding estimates~\citep[e.g.][]{Adams2012}. We report confirmed common proper motion companions in Figure~\ref{fig:astr_plot} and Tables~\ref{tab:comp_astr} and \ref{tab:prev_studies}. The two background objects are reported separately in Figure~\ref{fig:astr_plot_bg} and Table~\ref{tab:astr_bg}. We discuss each system individually in the Section~\ref{sec:companions} below.

\subsection{Masses and separation}
For each confirmed companion-primary pair, we compute a flux ratio based on the stars' physical parameters,
\begin{equation}
F_r=\frac{\int_{0}^{\infty} I(\lambda,T_{p},\log g_{p})\,R(\lambda)\,r_{p}^2\,\mathrm{d}\lambda}{\int_{0}^{\infty} I(\lambda,T_{c},\log g_{c})\,R(\lambda)\,r_{c}^2\,\mathrm{d}\lambda},
\label{eqn:flux}
\end{equation}
where the subscripts $p$ and $c$ refer to the primary and companion stars, respectively, $I(\lambda,T,\log g)$ is the star's specific intensity, $T$ is the star's effective temperature, $\log g$ is a measure of the stellar surface gravity, $R(\lambda)$ is the response function of the NIRC2 filter, and $r$ is the stellar radius. In order to determine $I(\lambda,T,\log g)$, we interpolate the gridded PHOENIX synthetic spectra~\citep{Husser2013} for solar metallicities and composition ([Fe/H]$=0$ and [$\alpha$/H]$=0$). Although our companion host stars have [Fe/H] between -0.21 and 0.25, we found that this assumption changes our estimated companion temperatures by less than $5\unit{K}$.

We use previously published measurements for the primary star's mass, radius, effective temperature and distances as listed in Table~\ref{tab:stellar_params} and fit for the companion star's effective temperature, $T_c$ by matching the observed flux ratio with the computed flux ratio from Equation~\ref{eqn:flux}. In order to calculate the flux from a companion with a given effective temperature, we use the zero-age main sequence models from \citet{Baraffe1998} to match each effective temperature with a corresponding radius and surface gravity, and then calculate the corresponding flux ratio using Equation~\ref{eqn:flux}. After determining the best-fit companion temperature, we calculate the corresponding uncertainty as the sum in quadrature of the uncertainties contributed by the flux ratio measurement error and the reported primary star temperature, surface gravity, and radius. We determine uncertainties for these parameters from uncertainties in our flux ratio measurement and primary star's temperature, radius, and $\log g$. We do not include any uncertainties introduced through use of the stellar model or the PHOENIX model spectra.

Finally, we convert our measured projected on-sky separations to projected spatial separations using the stellar distance. Unfortunately most of our stars don't have measured parallaxes; for these systems we generally used a spectroscopic distance estimate based on the derived stellar properties combined with the star's apparent magnitude.  

We calculate an estimated temperature for each candidate companion using either the J, H, or K band photometry, and find that these three independent temperature estimates are consistent with a late-type main sequence star in each filter.  This indicates that all of the detected companions have infrared colors consistent with their inferred effective temperatures. In Table~\ref{tab:interp_sec}, we report the error-weighted averages of the companion stellar parameters from all three bands as well as temperature estimates using data from individual bands. We conclude that our detected companion stars have colors consistent with a late-type main sequence star in the same system rather than being a distant early-type star in the background.

\subsection{Contrast curves}
\begin{figure}
\epsscale{1.1}
\plotone{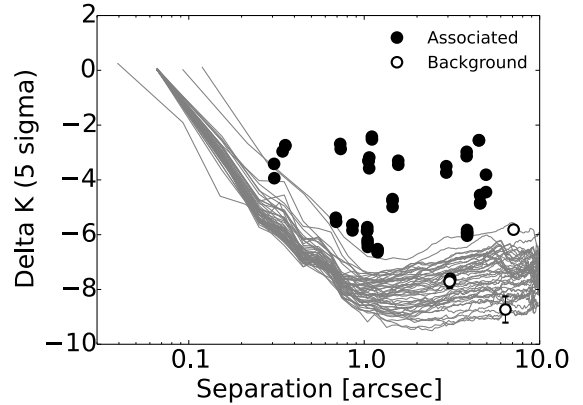}
\caption{The grey lines show K band contrast curve for all 50 Friends of Hot Jupiters AO targets are plotted. For each target, the curve with the greatest contrast was chosen. All companion stars are masked out for the computation of these curves. The points show the magnitude difference and separation for all detected candidate companions at all epochs, where filled points indicate physically associated companions (with confirmed common proper motion) and open points indicate background objects.
\label{fig:contrast}}
\end{figure}

We calculate contrast curves for our target stars as follows.  First, we measure the FWHM of the central star's PSF in the stacked and combined image, taking the average of the FWHM in the $x$ and $y$ directions as our reference value.  We then create a box with dimensions equal to the FWHM and step it across the array, calculating the total flux from the pixels within the box at a given position.  The $1\sigma$ contrast limit is then defined as the standard deviation of the total flux values for boxes located within an annulus with a width equal to twice the FWHM centered at the desired radial separation.  We convert our absolute flux limits to differential magnitude units by taking the total flux in a box of the same size centered on the peak of the stellar point spread function and calculating the corresponding differential magnitude at each radial distance.  We show the resulting $5\sigma$ average contrast curve for these observations in Figure~\ref{fig:contrast}.

\section{Systems with detected companions}
\label{sec:companions}

\subsection{New physically bound companions}
Seven out of our 17 targets with detected companions (HAT-P-10, HAT-P-14, HAT-P-16, HAT-P-24, HAT-P-33, WASP-3, and WASP-14) have not been previously reported to have a directly imaged stellar companion. Here, we discuss each of these systems individually and report error-weighted averages of corresponding measurements in Tables~\ref{tab:comp_astr} and~\ref{tab:interp_sec}.

\subsubsection{HAT-P-10}
HAT-P-10 hosts a transiting gas giant planet with a mass of $0.5\unit{\Mjup}$ and an orbital semimajor axis of of $0.04\unit{AU}$~\citep{Bakos2009a}. HAT-P-10b does not have a published Rossiter-McLaughlin measurement constraining its spin-orbit measurement, but its eccentricity is consistent with zero~\citep{Bakos2009a,West2009}. We therefore placed this system in our control sample. We report a $0.36\unit{\Msun} \pm 0.05\unit{\Msun}$ stellar companion at a projected separation of $42\unit{AU} \pm 2\unit{AU}$. In the first paper of the Friends of Hot Jupiter survey~\citep{Knutson2014}, HAT-P-10 was found to have a detected radial velocity trend that is consistent with the mass and separation of the directly imaged companion reported in this work. We show that HAT-P-10B is a common proper motion companion. Although there are three other systems with both radial velocity trends and a directly imaged stellar companion, this is the only such target where the stellar companion might plausibly explain the observed trend~\citep{Knutson2014}.

\subsubsection{HAT-P-14}
\citet{Torres2010} report a transiting gas giant planet around HAT-P-14 with a mass of $2.2\unit{\Mjup}$ and an orbital semimajor axis of $0.06\unit{AU}$. We placed this system in our misaligned sample because this planet has an eccentricity of $0.12 \pm 0.02$~\citep{Knutson2014} and a spin-orbit angle of $-171\deg\pm5\deg$~\citep{Winn2011}. We find a $0.20\unit{\Msun}\pm0.04\unit{\Msun}$ stellar companion to HAT-P-14 at a projected separation of $180\unit{AU} \pm 10\unit{AU}$. Our observations in 2012 through 2014 confirm that HAT-P-14B is a common proper motion companion. 

\subsubsection{HAT-P-16}
HAT-P-16 hosts a transiting gas giant planet with a mass of $4.2\unit{\Mjup}$ and an orbital semimajor axis of $0.04\unit{AU}$~\citep{Buchhave2010}. \citet{Moutou2011} measured a spin-orbit angle consistent with zero but \citet{Knutson2014} measures a planet orbital eccentricity of $0.04 \pm 0.01$ so we placed this system in our misaligned sample. We report a $0.19\unit{\Msun} \pm 0.03\unit{\Msun}$ stellar companion to HAT-P-16 at a separation of $160\unit{AU} \pm 10\unit{AU}$. From our detections in February 2012 and August 2013, we are able to confirm that HAT-P-16B is a common proper motion companion.

\subsubsection{HAT-P-24}
\citet{Kipping2010} report a transiting gas giant planet around HAT-P-24 with a mass of $0.7\unit{\Mjup}$ and an orbital semimajor axis of $0.05\unit{AU}$. They also find the planet's eccentricity to be consistent with zero and \citet{Albrecht2012} find evidence for a well-aligned orbit, so we placed this target in our control sample. We report a $0.33\unit{\Msun} \pm 0.03\unit{\Msun}$ stellar companion to HAT-P-24 at a separation of $2000\unit{AU} \pm 100\unit{AU}$. From our detections in February 2012 and May 2013, we are able to confirm that HAT-P-24B is a common proper motion companion.

\subsubsection{HAT-P-33}
HAT-P-33 is a planetary system with a transiting gas giant planet with a mass of $0.8\unit{\Mjup}$ and an orbital semimajor axis of $0.05\unit{AU}$~\citep{Hartman2011c}. We placed this system in our control sample because HAT-P-33b has an eccentricity consistent with zero~\citep{Hartman2011c} and no published spin-orbit angle measurement. We report a $0.55\unit{\Msun} \pm 0.03 \unit{\Msun}$ stellar companion to HAT-P-33 at a separation of $119\unit{AU} \pm 3\unit{AU}$. From our detections in February 2012 and March 2013, we are able to confirm that HAT-P-33B is a common proper motion companion.

\subsubsection{WASP-3}
WASP-3 hosts a transiting gas giant planet with a mass of $1.8\unit{\Mjup}$ and an orbital semimajor axis of $0.03\unit{AU}$~\citep{Pollacco2008}. WASP-3b has an eccentricity consistent with zero~\citep{Pollacco2008,Gibson2008} and a measured spin-orbit angle consistent with zero~\citep{Tripathi2010} so we placed this target in our control sample. We report a $0.108\unit{\Msun} \pm 0.006\unit{\Msun}$ stellar companion to WASP-3 at a separation of $300\unit{AU} \pm 20\unit{AU}$. From our detections in June 2012, July 2012, and May 2013, we are able to confirm that WASP-3B is a common proper motion companion.

\subsubsection{WASP-14}
\citet{Joshi2009} report a transiting gas giant planet around WASP-14 with a mass of  $7.3\unit{\Mjup}$ and an orbital semimajor axis of  $0.04\unit{AU}$. This system is in our misaligned sample because the planet has an eccentricity of  $0.082 \pm 0.003$~\citep{Knutson2014} and a spin-orbit angle of  $-33\deg\pm7\deg$~\citep{Johnson2009b}. We report a  $0.33\unit{\Msun} \pm 0.04\unit{\Msun}$ stellar companion to WASP-14 at a separation of  $300\unit{AU} \pm 20\unit{AU}$. From our detections in July 2012 and March 2013, we are able to confirm that WASP-14B is a common proper motion companion.

\subsection{Companions reported in previous studies}
\label{sec:prev_studies}
The remaining ten stellar companions were detected in previous studies, although not all systems were confirmed to be physically bound. We discuss each system individually below, with measurements reported in Table~\ref{tab:prev_studies}.

\subsubsection{HAT-P-7}
\label{sec:HATP7}
HAT-P-7 hosts a transiting gas giant planet with a mass of  $1.8\unit{\Mjup}$ and an orbital semimajor axis of  $0.04\unit{AU}$~\citep{Pal2008}. HAT-P-7b has a measured eccentricity consistent with zero~\citep{Pal2008} but its spin orbit angle is  $160\deg\pm40\deg$~\citep{Albrecht2012,Winn2009} so we placed this target in our misaligned sample. Our reported stellar companion to HAT-P-7 is consistent with the seven previous detections by \citet{Faedi2013,Narita2010b,Bergfors2013,Narita2012} between 2009 and 2012. The study by \citet{Narita2012} determined this detected stellar companion HAT-P-7B is common proper motion companion. Our own detections in 2012 through 2014 are also consistent with this conclusion. In addition, \citet{Narita2012} also reported a second directly imaged stellar companion (a ``western companion'') in four of their images (spanning the same dates as above) with a separation of $3.1\arcsec\pm0.1\arcsec$ and a position angle of $265.9\deg\pm0.4\deg$. They were unable to confirm that this object is bound to the primary. We were able image this second companion on three out of the four nights we observed HAT-P-7. Our analysis (Figure~\ref{fig:astr_plot_bg}) concludes that this second companion is actually a background object and therefore not physically bound to the other two stars. This target is one of four systems with both a directly imaged stellar companion and a long term radial velocity trend, as reported in \citet{Knutson2014}. However, our resolved stellar companion,  at a projected separation of $1240\unit{AU}\pm70\unit{AU}$, is too distant to explain the observed radial velocity trend.

\subsubsection{HAT-P-8}
\citet{Mancini2013} report a transiting gas giant planet around HAT-P-8 with a mass of  $1.3\unit{\Mjup}$ and an orbital semimajor axis of  $0.04\unit{AU}$. They also report the planet's eccentricity to be consistent with zero and its spin-orbit angle has been measured to be consistent with zero~\citep{Simpson2011,Moutou2011}. Thus, we placed this system in our control sample. In our survey, we find two stellar companions for HAT-P-8. These companions were previously detected by \citet{Bergfors2013} in 2009 but their photometry did not resolve the two stars and they instead reported a single companion. We resolve these two stars in our Keck images, which were published in \citet{Bechter2014}. \citet{Bechter2014} followed up on this target in 2012 and 2013 and find that both stellar companions has common proper motion with the primary. In this work we present a new set of observations, also taken in 2012 and 2013, which also indicate that HAT-P-8 is a physically bound triple system. We do not use the unresolved astrometric measurements from \citet{Bergfors2013} in our analysis because the resolved measurements from \citet{Bechter2014} and this survey were enough to confirm common proper motion.

\subsubsection{HAT-P-30}
HAT-P-30 is a planetary system with a transiting gas giant planet with a mass of  $0.7\unit{\Mjup}$ and an orbital semimajor axis of  $0.04\unit{AU}$~\citep{Johnson2011}. Although they measure the planet's eccentricity to be consistent with zero, they also measure a spin-orbit misalignment of  $74\deg\pm9\deg$. We therefore placed this target in our misaligned sample. Our detected stellar companion is consistent with the companion reported by \citet{Adams2013}, observed between 2011 Oct 9-11. With only one detection, they were unable to confirm common proper motion. We found this companion in 2012 and followed up in 2014. Our data confirm that the companion HAT-P-30B is indeed bound. We do not use the~\citet{Adams2013} measurement in this analysis as they did not report uncertainties on their measured separation and position angle.

\subsubsection{HAT-P-32}
HAT-P-32 hosts a transiting gas giant planet with a mass of  $0.9\unit{\Mjup}$ and an orbital semimajor axis of  $0.03\unit{AU}$~\citep{Hartman2011c}. HAT-P-32b has an eccentricity consistent with zero~\citep{Hartman2011c} but a mesaured spin-orbit misalignment of  $85\deg\pm2\deg$~\citep{Albrecht2012} so we placed this system in our misaligned sample. Our detected stellar companion is consistent with the stellar companion reported by \citet{Adams2013}, observed between 2011 Oct 9-11. As with HAT-P-30, they were unable to measure the proper motion for this companion. We found this stellar companion in 2012 and followed up in 2013. Our data confirms that HAT-P-32B is indeed bound. We do not use the~\citet{Adams2013} measurement in this analysis as they did not report uncertainties in the measured separation and position angle. This target is also one of four targets with both a directly imaged stellar companion and a long term radial velocity trend, as reported in \citet{Knutson2014}. However, our resolved stellar companion,  at a projected separation of $830\unit{AU}\pm20\unit{AU}$, is too distant to explain the observed radial velocity trend.

\subsubsection{TrES-2}
\citet{Barclay2012} report a transiting gas giant planet around TrES-2 with a mass of  $1.4\unit{\Mjup}$ and an orbital semimajor axis of  $0.04\unit{AU}$. They also report a planet eccentricity consistent with zero and \citet{Winn2008a} report a spin-orbit angle consistent with zero so we placed this target in our control sample. Our detected stellar companion for TrES-2 is consistent with the three previous detections reported by \citet{Daemgen2009}, \citet{Faedi2013}, and \citet{Bergfors2013} between 2007 and 2009. These studies were not able to measure the proper motion of the companion in order to determine whether or not it was bound to TrES-2. Our measurements from 2012 and 2013 show that TrES-2B is a common proper motion companion.

\subsubsection{TrES-4}
TrES-4 is a planetary system with a transiting gas giant planet with a mass of  $0.9\unit{\Mjup}$ and an orbital semimajor axis of  $0.05\unit{AU}$~\citep{Sozzetti2009}. TrES-4b has an eccentricity consistent with zero~\citep{Sozzetti2009} and \citet{Narita2010a} measure a spin-orbit angle consistent with zero so we placed this system in our control sample. Our detected stellar companion for TrES-4 is consistent with the three previous detections reported by \citet{Daemgen2009}, \citet{Faedi2013}, and \citet{Bergfors2013} between 2008 and 2009. Using all the data from the prior studies, \citet{Bergfors2013} were able to confirm that TrES-4B is a common proper motion companion. Our measurements from 2012 and 2013 support this assessment.

\subsubsection{WASP-1}
WASP-1 hosts a transiting gas giant planet with a mass of  $0.8\unit{\Mjup}$~\citep{Knutson2014} and an orbital semimajor axis of  $0.04\unit{AU}$~\citep{Torres2008}. \cite{Albrecht2011} report that the eccentricity and spin-orbit angles are consistent with zero, so we placed this target in our control sample. \citet{CollierCameron2007} observed a stellar companion to WASP-1 with a separation of $4.7\arcsec$ north of the primary from observations made in 2006. They did not report any uncertainties nor did they provide a numerical positional angle. Thus, we are unable to include this observation in our analysis. However, these values are consistent with the reported separation and position angle from our observations in 2012, 2013 and 2014. Our measurements confirm that WASP-1B is a common proper motion companion.

\subsubsection{WASP-2}
There is a transiting gas giant planet around WASP-2 with a mass of  $0.9\unit{\Mjup}$~\citep{Knutson2014} and an orbital semimajor axis of  $0.06\unit{AU}$~\citep{Torres2008}. Radial velocity and secondary eclipse measurements for this planet are consistent with a circular orbit, but \citet{Triaud2010} find that the planet has spin-orbit misalignment of  $150{\deg}^{+10\deg}_{-20\deg}$.  We therefore place this planet in the misaligned sample. A stellar companion for WASP-2 was reported in six separate observations from 2006 to 2011 by \citet{CollierCameron2007}, \citet{Daemgen2009}, \citet{Bergfors2013}, and \citet{Adams2013}. These observations are all consistent with the companion we detect in our images. However, the measurements reported in 2006 by \citet{CollierCameron2007} did not include uncertainties nor a numerical value for the position angle, so we do not use this observation in our analysis. In addition, one observation from 2011 by \citet{Adams2013} also did not include uncertainties so we omit this measurement as well. \citet{Bergfors2013} combined their observations with previous studies to conclude that the stellar companion must be bound to the primary. Our own measurements from 2012 and 2013 also indicate WASP-2B is a common proper motion companion.

\subsubsection{WASP-8}
WASP-8 is a planetary system with a transiting gas giant planet with a mass of  $2.2\unit{\Mjup}$ and an orbital semimajor axis of  $0.08\unit{AU}$~\citep{Queloz2010}. \citet{Knutson2014} reports the planet's eccentricity to be  $0.304 \pm 0.004$ and the discovery paper measures a spin-orbit misalignment of  $-123{\deg}^{+3\deg}_{-4\deg}$. Therefore, we placed this planet in the misaligned sample. This paper also reported a stellar companion for WASP-8 consistent with our own observations in 2012 and 2013. They did not provide a date for their observation, so we do not use their measurement in our analysis. However, they note that this companion was mentioned in the Washington Double Star Catalog~\citep{Mason2001} from 70 years ago at the same position. They conclude that WASP-8B is a common proper motion companion. Our data also rules out a background object and supports the conclusion of \citet{Queloz2010}. This target is one of four targets with a directly imaged companion and also a long term radial velocity trend, as reported in \citet{Knutson2014}. However, our resolved stellar companion,  at a projected separation of $380\unit{AU}\pm50\unit{AU}$, is too distant to explain the observed radial velocity trend.

\subsubsection{WASP-12}
WASP-12 hosts a transiting gas giant planet with a mass of  $1.4\unit{\Mjup}$ and an orbital semimajor axis of  $0.02\unit{AU}$~\citep{Hebb2009}. WASP-12b has eccentricity consistent with zero~\citep{Hebb2009}, however, \citet{Albrecht2012} measures a spin-orbit angle of  $59{\deg}^{+15\deg}_{-20\deg}$. Thus, we placed this system in our misaligned sample. Stellar companion(s) for WASP-12 have been seen in 7 observations from 2009 to 2013 as reported by~\citet{Bergfors2013}, \citet{Crossfield2012}, \citet{Sing2013}, and \citet{Bechter2014}. The first two studies were not able to resolve the two stellar companions for WASP-12 but did detect a single source at the same position. \citet{Bergfors2013} noted that the source was elongated and \citet{Crossfield2012} obtained a spectrum for the companion, which they used to estimated its effective temperature and surface gravity.  They concluded that the observed source must be a foreground object as it was otherwise significantly brighter than expected for its spectral type. \citet{Sing2013} was the first to report images of the two resolved stellar companions but did not provide any astrometric measurements so we do not include this data point in our analysis. The two observations reported in \citet{Bechter2014} are part of the Friends of Hot Jupiters survey. We independently analyzed the data and our results are consistent with the measurements reported in \citet{Bechter2014}. Our measurement uncertainties are lower as we account for the NIRC2 distortion; thus we use the measurements reported in this work for our analysis. We also recently discovered a more precise distance estimate for this star in \citet{Triaud2014}, which we use to replace the \citet{Bergfors2013} value from our previous papers. Despite these improvements, we find that our 2012 and 2013 data points are insufficient to rule out the possibility of a background source (see astrometric plots for WASP-12B and WASP-12C separately in Figure~\ref{fig:astr_plot}; these plots do not include the combined light astrometric measurements).  However, when considering previously reported combined light measurements for this system our analysis indicate the center of mass of WASP-12BC has common proper motion with the primary, consistent with our conclusions in \citet{Bechter2014}.

\subsection{Candidates determined to be background objects}
As discussed in Section~\ref{sec:HATP7}, we found that the second candidate companion to HAT-P-7 reported in \citet{Narita2012} and followed up in our survey was a background object (see Figure~\ref{fig:astr_plot_bg} and Table~\ref{tab:astr_bg}), so HAT-P-7 is simply a binary system.

We also detect a candidate companion to HAT-P-15 on 2 February 2012 with $\Delta \Kpfilt = 5.77\pm0.05$, a separation of $7.100\arcsec \pm 0.002\arcsec$ and position angle of $58.6\deg \pm 0.1\deg$. Due to the large separation of this object to HAT-P-15, the object was only visible in one of our dither positions in our first epoch, resulting in relatively large astrometric uncertainties. We followed up with a second epoch on 3 October 2014 and did not find an object in the same position. However, we did find an object with $\Delta \Ksfilt = 8.2\pm0.9$, a separation of $6.387\arcsec \pm 0.008\arcsec$ and position angle of $58.0\deg \pm 0.1\deg$. Physical association of these two detections are ruled out by our astrometric analysis (see Figure~\ref{fig:astr_plot_bg} and Table~\ref{tab:astr_bg}). In addition, the location of these detections are also inconsistent with both detections being the same object. Thus, we conclude that we imaged two different background or foreground objects in the two epochs and exclude this target from our list of bound companions.

\subsection{Non-detections}
For the remaining 32 targets, we did not find any candidate companions within the 5.5\arcsec\ NIRC2 field of view. For two of these targets, previous studies have found directly imaged companions at larger separations. \citet{Mugrauer2014} reports a common proper motion companion to the south of HAT-P-4 with a separation of $91.8\arcsec$. \citet{Bergfors2013} reports a candidate companion to XO-3 with a separation of $6.059\arcsec\pm0.047\arcsec$ and a position angle of $296.7\deg\pm0.3\deg$. This companion is also very faint ($\Delta z^\prime=8.22\pm0.23$ and $\Delta i^\prime=8.57\pm0.24$) and \citet{Bergfors2013} note that it is unlikely that the detected object is a bound companion.  This companion would not be visible in our survey due to its faintness and our survey's FOV, given our 3-point dither pattern.

\section{Companion Fraction}
\label{sec:compfrac}
\begin{figure}
\epsscale{1.2}
\plotone{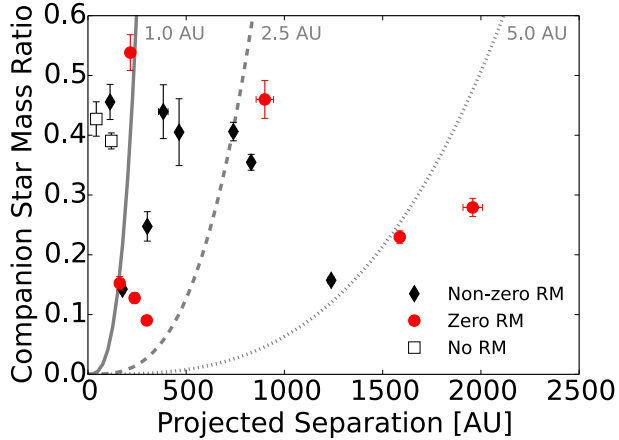}
\caption{For each target, we plot the weighted average of companion masses and projected separation from all epochs reported in Table~\ref{tab:interp_sec}. Only one point is plotted for each of the triple systems. We plot targets with a non-zero spin-orbit angle measured using the Rossiter-McLaughlin (RM) effect as black diamonds, targets with spin-orbit angles consistent with zero as red circles, and targets without any spin-orbit angle measurement as open squares. The two objects without a RM measurement are HAT-P-10b and HAT-P-33b; both have orbital eccentricities consistent with zero so we placed them in the ``control'' sample. Companions that lie above and to the left of the solid, dashed, and dotted lines are sufficiently massive and close enough to excite Kozai-Lidov oscillations and overcome general relativity pericenter precession for giant planets at $1\unit{AU}$, $2.5\unit{AU}$ and $5\unit{AU}$, respectively.
\label{fig:sep_v_mass}}
\end{figure}

We find bound stellar companions around 17 out of the 50 targets observed, corresponding to an overall raw companion fraction of $34\%\pm7\%$. We find that 9 out of 27 stars with planets on misaligned or eccentric orbits have stellar companions, yielding a raw stellar companion fraction of $33\%\pm9\%$. We also find that 8 out of 23 stars with planets on well-aligned or circular orbits have stellar companions, corresponding to a raw stellar companion fraction of $35\%\pm9\%$. Figure~\ref{fig:sep_v_mass} plots the physical separation of detected companions versus their mass ratios. Companions to stars with measured spin-orbit misalignment are shown in black, companions to stars with measured spin-orbit angles consistent with zero are shown in red, and companions to stars with no spin-orbit angle measurement (HAT-P-10 and HAT-P-33) are shown as open squares. We find no evidence for a difference in the typical mass ratios or projected separations between misaligned and well-aligned targets.

Figure~\ref{fig:sep_v_mass} also plots lines to show the minimum companion mass necessary to excite Kozai-Lidov oscillations at a timescale short enough to overcome pericenter precession due to general relativity. For these representative lines, we use a central stellar mass of $1\unit{\Msun}\,$, a planetary mass of $1\unit{\Mjup}\,$, and assume a circular orbit for both planet and companion star. We plot these limits for initial planet semimajor axis distances of $1\unit{AU}$, $2.5\unit{AU}$ and $5\unit{AU}$ (solid, dashed and dotted lines, respectively). These lines are computed by equating the timescales for the Kozai-Lidov oscillation and general relativity pericenter precession, using Equations 1 and 23 from~\citet{Fabrycky2007}. These expressions scale as $1-e^2$, so they are not strongly affected by our assumption of circular orbits. All companions that lie above and to the left of these lines are sufficiently massive and close enough to excite Kozai-Lidov oscillations on the planet. We find that the majority of our detected companions could potentially excite Jupiter-mass planets that form within $5\unit{AU}$.

\subsection{Survey incompleteness correction}
In order to make a reliable estimate of the companion fraction for our sample, we must correct for our survey incompleteness. For each of our 50 targets, we determine our survey's sensitivity to companions with various mass ratios and orbital semimajor axes. We create a 50x50 grid of linearly spaced bins in mass ratio and logarithmically spaced bins in semimajor axis out to the distance that corresponds to a separation of 5.5\arcsec. For each grid point, we generate 1000 fake companions with a mass and semimajor axis within the grid point limits. We draw the companion's orbital eccentricity from a uniform distribution derived from surveys of field star binaries~\citep{Raghavan2010} and we randomize the remaining orbital elements. Then, for each fake companion, we compute the projected angular separation and brightness ratio and compare to that target's $5\sigma$ contrast curve to determine whether or not our survey would have been able to detect that fake companion. Thus, we compute our survey sensitivity for star $i$ for companions in the grid point corresponding to mass ratio $m_r$ and semimajor axis $a$ to be $D_i(m_r,a)$.

We then compute the total average sensitivity, or survey completeness, for star $i$ as $S_i$ by taking the average of all $D_i(m_r,a)$ values weighted by the frequency of a companion with each $m_r$ and $a$: $f(m_r,a)$. \citet{Raghavan2010} determines $f(m_r,a)$ to be a log-normal distribution in period (which is a function of $m_r$ and $a$). We use this distribution to compute each target star's survey sensitivity as:
\begin{equation}
S_i = \frac{\int\int D_i(m_r,a)f(m_r,a) \mathrm{d} m_r \mathrm{d} \ln a}{\int\int f(m_r,a)\mathrm{d} m_r \mathrm{d} \ln a}
\label{eqn:sensitivity}
\end{equation}
where we integrate over the range of our survey. We are sensitive to companions with periods between $10^4\unit{days}$ and $10^{7.5}\unit{days}$, which approximately corresponds to semimajor axes between $50\unit{AU}$ and $2000\unit{AU}$.

We use our estimate of survey completeness for each star to compute the likelihood $L$ of obtaining our data set of $N_d$ detected companions out of $N$ survey targets as
\begin{equation}
L = \prod_{i=1}^{N_d} (S_i \eta)  \prod_{j=1}^{N-N_d}(1-S_j\eta)
\label{eqn:likelihood}
\end{equation}
where $\eta$ is the true companion fraction and the product sum over $i$ is for the targets with a detected companion while the product sum over $j$ is for the targets without a detected companion. We define companion fraction as the fraction of stars with at least one stellar companion within our survey range.

We use the Affine-Invariant Markov Chain Monte Carlo scheme implemented by the python package ``emcee''~\citep{Goodman2010,ForemanMackey2013} to determine the posterior probability distribution of $\eta$. Our prior on $\eta$ is uniform between $\eta=0$ and $\eta=1$. In addition, we assume that we are 100\% complete for targets where we have detected at least one companion because we expect all triple or higher order systems to be hierarchical. That is, we set all $S_i=1.0$ but still compute $S_j$ as described in Equation~\ref{eqn:sensitivity}. This assumption is supported by the two hierarchical systems detected by our survey and by previous studies~\citep[e.g.][]{Eggleton2007}.

\begin{figure}
\epsscale{1.2}
\plotone{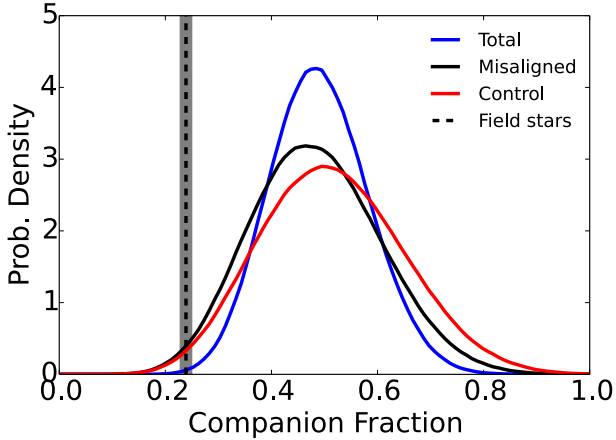}
\caption{A comparison of the posterior probability density of the fraction of stars with at least one companion, $\eta$, for our total sample, our misaligned sample, our control sample, and for solar type field stars in the solar neighborhood~\citep{Raghavan2010}. Only companions with periods between $10^4\unit{days}$ and $10^{7.5}\unit{days}$ are considered.
\label{fig:posterior_eta}}
\end{figure}

We find that the companion fraction is $\eta_T=49\%\pm9\%$ for our total sample, $\eta_M=48\%\pm12\%$ for our misaligned sample, and $\eta_C=51\% \pm 13\%$ for our control sample. These posterior distributions are shown in Figure~\ref{fig:posterior_eta}. These fractions are consistent with one another and we therefore conclude that there is no evidence for a correlation between the presence of a stellar companion and the orbital properties of the transiting gas giant planet.

We compare the companion fraction for hot ($T_{\mathrm{eff}}>6200\unit{K}$) and cool primary stars and find that hot stars have a higher companion rate, at the $2.9\,\sigma$ level. We find the companion fraction for hot stars to be $75\%\pm14\%$ and the companion fraction for cool stars to be $34\%\pm10\%$. This difference is consistent with surveys of stellar multiplicity, which indicate that more massive stars have higher binary fractions~\citep[e.g.][]{Duchene2013}. Surveys of hot Jupiter obliquities indicate that misaligned hot Jupiters are preferentially located around stars hotter than $6200\unit{K}$, which may be due to more efficient tidal damping of primordial spin-orbit misalignments in systems with cooler host stars \citep{Schlaufman2010,Winn2010b,Albrecht2012}. We test this hypothesis by recalculating the companion fractions for our two sub-samples using only stars hotter than $6200\unit{K}$, where planets should still retain their primordial spin-orbit orientations. We find that hot stars in our misaligned sample have a companion fraction of $59\%\pm17\%$, while hot stars in our control sample have a companion fraction of $83\%\pm14\%$. These fractions are consistent at the $1.7\,\sigma$ level.

We also consider the companion fractions for the 35 stars in our sample with published measurements of spin-orbit alignment from the Rossiter-McLaughlin effect. We find the companion fraction for stars with a non-zero spin-orbit angle to be $73\%\pm15\%$ while the companion fraction for stars with a spin-orbit angle consistent with zero is $53\%\pm14\%$. These fractions are consistent at the $1.4\,\sigma$ level.

When we examined these same targets using the radial velocity technique we found that $51\% \pm 10\%$ had companions with masses between $1-13\,\Mjup$ on long period orbits~\citep{Knutson2014}. Aside from HAT-P-10, where the directly imaged stellar companion is consistent with the measured RV acceleration, we find no evidence for any correlation between the presence of a stellar companion and a measured RV acceleration.  Approximately 1/3 of our target stars have a detected stellar companion and 1/3 have a radial velocity acceleration, so we would therefore expect to see approximately one-ninth of the stars in our sample with both types of companions.  We detect four such systems including HAT-P-10.  We therefore conclude that the rates of stellar and radial velocity companions appear to be independent of one another and sum the two in quadrature to obtain a combined stellar and planetary occurrence rate of $72\% \pm 16\%$.

\subsection{Comparison with other direct imaging surveys}
We compare this companion fraction to the results from other surveys for stellar companions to planet hosts and field stars. These surveys are summarized in Table~\ref{tab:other_surveys}. Our survey results are consistent with previous direct imaging surveys for stellar companions to transiting gas giant planet hosts~\citep{Daemgen2009,Faedi2013,Bergfors2013,Adams2013}. These previous surveys had small sample sizes; the largest sample size was 21 stars~\citep{Bergfors2013}. Our results are also consistent with the result of infrared direct imaging surveys of {\it Kepler} transiting planet candidate hosts~\citep{Adams2012,Adams2013,Dressing2014,LilloBox2012,LilloBox2014,Wang2014}. Except for \citet{Wang2014} and \citet{Horch2014}, none of the above surveys quantify their completeness and their numbers are closer to our uncorrected companion fraction. In addition, these surveys have quite widely varying sensitivities that may contribute to the large observed scatter in measured companion fraction.

\begin{figure}
\epsscale{1.2}
\plotone{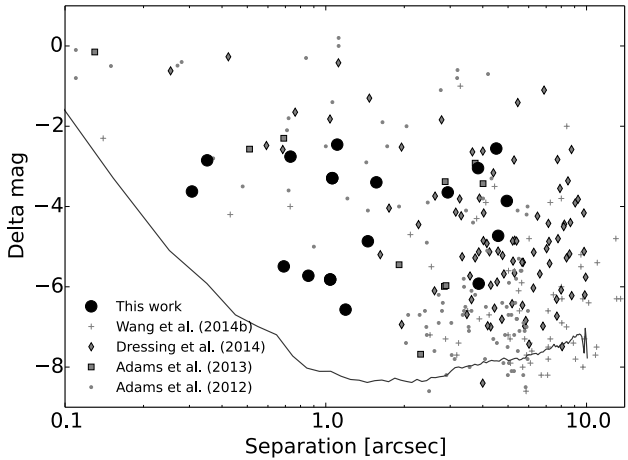}
\caption{The difference in magnitude (\Jfilt\ or \Ksfilt) for our confirmed physically associated companions compared to companion candidates from other near-infrared direct imaging surveys and a representative $5\,\sigma$ contrast curve from our survey. Our survey is more sensitive to faint companions at small separations. The published surveys shown here target {\it Kepler} planet candidate host stars, which are on average three times more distant than our targets. Thus, our survey finds companions at smaller projected physical separations as well.
\label{fig:sep_v_dmag}}
\end{figure}

Figure~\ref{fig:sep_v_dmag} compares the companions reported in this work with these diffraction-limited near-infrared AO surveys. As companion masses are not always reported, we plot the difference in magnitude versus angular separation for all reported companions. Our survey is primarily sensitive to companions within 5\arcsec\ while the others consider companions at larger separations. In addition, our survey is able to detect companions that are approximately 2 magnitudes fainter than these previous studies at separations less than 2\arcsec. Because the {\it Kepler} candidate host stars are on average significantly more distant than our target stars, these surveys were unable to distinguish between background objects and bound stellar companions. This also means that the {\it Kepler} surveys are less sensitive to stellar companions at small projected physical separations. We also limit our survey to systems with short-period transiting gas giant planets, while the {\it Kepler} planet candidate sample is dominated by much smaller planets, many of which are in compact multi-planet systems.

The direct imaging surveys conducted at visible wavelengths~\citep{Law2014,Gilliland2015,Horch2014} report smaller companion fractions than our survey and other NIR surveys. This is not surprising, as the majority of our detected companions are M stars that would be much fainter at optical wavelengths. These surveys also have relatively small fields of view, which would also contribute to a lower detection rate for companions as compared to the wide-field infrared surveys.

We note that our results for the multiplicity rate of hot Jupiter host stars may also have implications for the ongoing debate about the origin of the discrepancy between hot Jupiter occurrence rates from radial velocity and transit surveys. \citet{Wang2014arxiv} point out that hot Jupiter occurrence rates in {\it Kepler} surveys are approximately two to three times smaller than hot Jupiter occurrence rates in Doppler planet surveys. Because the {\it Kepler} survey does not filter multi-stellar systems from its target list, \citet{Wang2014arxiv} suggest that one potential cause for the discrepancy may be that these companion stars are suppressing the formation of all planets, including hot Jupiters, in the {\it Kepler} sample. \citet{Wang2014} also finds that planets are $1.7\pm0.5$ times less likely to form in a system with a stellar companion within $1000\unit{AU}$. However, we find that half of all hot Jupiters are found in stellar binaries, indicating that stellar multiplicity does not inhibit the formation of these systems for cases where the stellar companion is on an orbit approximately 50-2000$\unit{AU}$.

\subsection{Comparison with field stars}
\label{sec:compare_field}
Finally, we compare our companion fraction to the measured multiplicity rate for Solar-type field stars. The most recent survey~\citep{Raghavan2010} reports that $44\% \pm 2\%$ of Solar-type stars within $25\unit{pc}$ are in multiple star systems, which is in good agreement with previous results from \citet{Duquennoy1991}. This value is corrected for survey completeness. This fraction includes companions with separations too small to resolve with direct imaging or too wide to be found within our survey's field of view. Thus, we only consider the companions from \citet{Raghavan2010} with periods between $10^{4}\unit{days}$ and $10^{7.5}\unit{days}$, which corresponds to physical separations of approximately $50\unit{AU}$ and $2000\unit{AU}$. If we select only the subset of binaries from \citet{Raghavan2010} that fall within this period range, we find that the fraction of field stars with stellar companions is $24\%\pm1\%$. This is $2.8\,\sigma$ smaller than our estimated companion frequency for systems with short period gas giant planets, suggesting that the presence of a stellar companion increases the likelihood that a gas giant planet will migrate inward to a relatively short period orbit.

\section{Summary}
\label{sec:summary}
We found nineteen companions around seventeen targets, including two triple systems, HAT-P-8 and WASP-12, which we previously reported in ~\citet{Bechter2014}. We measure the proper motions for all detected companions and confirm that they are physically bound to the planet-hosting primary stars. We report seven new multiple star systems with transiting giant planets and provide follow-up observations for ten previously reported candidate multi-star systems. Our follow-up observations allowed us to confirm the bound nature of three previously detected candidate companions and were in good agreement with the conclusions of previous studies on the bound nature of the other seven companions. We also determined the second candidate companion to HAT-P-7 found by \citet{Narita2012} to be a background object. For all systems, we provide updated astrometric measurements as well as estimated masses and physical separations for the observed companions. We find that most of the detected companions are massive enough to excite Kozai-Lidov oscillations on giant planets forming within $5\unit{AU}$.

Our companion fraction is consistent with previous NIR direct imaging surveys of stellar companions around transiting planet hosts. We correct for survey sensitivity and find that close-in transiting gas giant planetary hosts are approximately twice as likely to have at least one stellar companion with $50\unit{AU}\lesssim a\lesssim2000\unit{AU}$ as compared to field stars, although the significance of this difference is only $2.8\,\sigma$. We find that the companion fraction for systems hosting a transiting gas giant planet on a misaligned or eccentric orbit is indistinguishable from the companion fraction for systems hosting a planet on a well-aligned and circular orbit. This is consistent with other results that suggest hot Jupiters may not primarily migrate via the Kozai-Lidov mechanism~\citep{Dawson2015,Naoz2012,Petrovich2015}. We also recalculate the companion fractions for our two sub-samples using only stars hotter than $6200\unit{K}$, as it has been suggested that tidal evolution might be able to remove primordial spin-orbit misalignments for planets orbiting cooler stars.  We find that there is no evidence for a correlation between the presence of a stellar companion and spin-orbit misalignment of the transiting hot Jupiter, in good agreement with our results for the full sample.  Finally, we calculate the companion frequency for the overall sample as a function of host star temperature and find that stars hotter than $6200\unit{K}$ have a companion rate that is approximately two times larger ($2.9\,\sigma$ significance) than their cool counterparts, consistent with surveys of stellar multiplicity~\citep{Duchene2013}.

We conclude that the data are consistent with two possible scenarios and discuss them here. First, stellar companions may not play a role in the determination of hot Jupiter spin-orbit misalignments. This would be consistent with the planet-planet scattering scenario~\citep[e.g.][]{Chatterjee2008,Nagasawa2008,Wu2011,Beauge2012,Lithwick2014}. However, our radial velocity survey in \citet{Knutson2014} found no correlation between the presence of a long term radial velocity acceleration and the spin-orbit alignment of the inner transiting planet. Other proposed orbital obliquity excitation mechanisms that do not require a companion star include misalignment of the natal disk's angular momentum vector with respect to the stellar spin axis due to chaotic star formation~\citep{Bate2010,Thies2011,Fielding2014}, magnetic warping torques due to the primary star's magnetic field~\citep{Lai2011}, and modulation of stellar surfaces by internal gravity waves~\citep{Rogers2012,Rogers2013b}. Alternatively, primordial gravitationally bound stellar companions may have acted to perturb the protoplanetary disks out of alignment with their host stars at the epoch of star and planet formation~\citep[e.g.][]{Batygin2012,Batygin2013,Crida2014,Storch2014}. However, dynamical processing by cluster evolution would have removed or exchanged these companions, diminishing their current observational signatures~\citep{Malmberg2007b}. In other words, the companions we observe today may not be the ones responsible for the facilitation of hot Jupiter misalignments. 

In this scenario, a majority of planetary systems should form from disks with random alignments, regardless of whether or not there is currently a stellar companion present. This prediction can be tested by measuring the spin-orbit alignment of a large number of coplanar, multi-planet systems such as those detected by the {\it Kepler} survey.  If a significant fraction of these systems are misaligned even when no stellar companion is present, it would provide strong support for the ubiquity of primordial disk misalignments, which could also explain the observed population of misaligned hot Jupiters. \citet{Morton2014b} present the framework for such a test.

Whatever the favored scenario may be, it must also be consistent with the increased frequency of misaligned hot Jupiters found around hot stars. If a stellar companion is not responsible for misalignment, then tidal evolution of the star-planet pair, which proceeds at an enhanced rate around cooler stars, could give rise to the observed trend~\citep[e.g.][]{Winn2010b,Lai2012,Valsecchi2014}. However, this mechanism has been recently criticized by \citet{Rogers2013a}, who argue that such a process requires an unphysical set of assumptions, and would generally lead to a misalignment distribution that is inconsistent with the observed one. Alternatively, magnetically facilitated disk-star coupling may cause cooler stars to realign with their disks~\citep{Spalding2014DPS}, signaling constancy with the second scenario where protoplanetary disks are misaligned by transient stellar companions at early times.

The apparent enhancement in the companion fraction for our sample of transiting planets versus that of field stars is suggestive, and may indicate that these companions play a role in the migration process. Intriguingly, \citet{Law2014} found tentative evidence that in the {\it Kepler} sample, short period gas giant planets are more likely to have stellar companions than their more distant counterparts, which is also consistent with the idea that these companions play a role in planet migration. In addition, when considering the companion fraction found in our radial velocity survey~\citep{Knutson2014}, we find that the overall rate of both planetary and stellar companions in systems with close-in transiting gas giant planets is $72\% \pm 16\%$, suggesting that these systems frequently have companions that may interact dynamically with the short-period planet.

In the future, we plan to survey a larger sample of planets detected using the radial velocity technique, which will span a much broader range of semimajor axes than any of the transiting planet surveys. A number of sources show long-term Doppler accelerations indicating the presence of outer companions, and Keck AO imaging has been demonstrated as a successful technique for identifying faint stellar and substellar companions based on the existence of such ``trends''~\citep{Crepp2012}. These data should provide a more definitive test of the potential correlation between multiplicity and orbital semimajor axis of the inner planet.  We also plan to expand our sample of hot Jupiter systems with AO imaging, in order to reduce the uncertainties in our estimate of the stellar companion rate for these systems.

\acknowledgments

This work was supported by NASA grant NNX14AD24G. HN is grateful for funding support from the Natural Sciences and Engineering Research Council of Canada. JAJ gratefully acknowledges support from generous fellowships from the David \& Lucile Packard and Alfred P. Sloan foundations. We also thank Brendan Bowler for assistance with our common proper motion analysis and Rebekah Dawson and Cristobal Petrovich for helpful discussions.

This work was based on observations at the W. M. Keck Observatory granted by the California Institute of Technology. We thank the observers who contributed to the measurements reported here and acknowledge the efforts of the Keck Observatory staff. We extend special thanks to those of Hawaiian ancestry on whose sacred mountain of Mauna Kea we are privileged to be guests.

\clearpage
\LongTables
\begin{deluxetable}{lclccccc}
\tablecolumns{8}
\tabletypesize{\scriptsize}
\tablecaption{Summary of NIRC2 AO Observations  \label{tab:obs}}
\tablewidth{0pt}
\tablehead{
\colhead{Target} & \colhead{$N_c$} & \colhead{UT Obs. Date} & \colhead{Filter} & \colhead{Array} & \colhead{$T_{\mathrm{int}}$} & \colhead{$N_{\mathrm{fit}}$} & \colhead{$N_{\mathrm{stack}}$}
}
\startdata
GJ-436 	& 0 & 2012 Feb 02	& \Kpfilt	& 256 & 5 & \nodata & 18\\

HAT-P-2	& 0 & 2012 May 29 & \Kpfilt	& 512 & 13.3 & \nodata & 9\\

HAT-P-4	& 0 & 2012 Feb 02	& \Kpfilt	& 1024 & 15 & \nodata & 9\\

HAT-P-6	& 0 & 2012 Jul 27	& \Kpfilt	& 1024 & 9 & \nodata & 30\\

HAT-P-7	& 1 & 2012 Jul 27	& \Jfilt	& 1024 & 9 & 10 & 10\\ 
& & 2012 Jul 27	& \Hfilt	& 1024 & 9 & 10 & 10\\
& & 2012 Jul 27	& \Kpfilt	& 1024 & 9 & 10 & 10\\
& & 2013 May 31	& \Ksfilt	& 1024 & 9 & 8 & 8\\ 
& & 2013 Jun 22	& \Hfilt	& 1024 & 9 & 12 & 12\\
& & 2013 Jun 22	& \Ksfilt	& 1024 & 9 & 12 & 12\\
& & 2014 Jul 12   & \Ksfilt	& 1024 & 15 & 15 & 15\\

HAT-P-8	& 2 & 2012 Jul 27	& \Jfilt	& 1024 & 9 & 10 & 15\\
& & 2012 Jul 27	& \Hfilt	& 1024 & 9 & 13 & 15\\
& & 2012 Jul 27	& \Kpfilt	& 1024 & 9 & 15 & 15\\
& & 2013 Aug 19	& \Hfilt	& 512 & 12.5 & 6 & 6\\
& & 2013 Aug 19	& \Kpfilt	& 1024 & 9 & 12 & 12\\
& & 2013 Aug 19	& \Ksfilt	& 1024 & 9 & 12 & 12\\

HAT-P-10	& 1 & 2012 Feb 02	& \Jfilt	& 1024 & 9 & 9 & 9\\
& & 2012 Feb 02	& \Kpfilt	& 1024 & 10 & 9 & 9\\
& & 2013 Aug 19	& \Hfilt	& 1024 & 9 & 6 & 6\\
& & 2013 Aug 19	& \Kpfilt	& 1024 & 9 & 12 & 12\\
& & 2013 Aug 19	& \Ksfilt	& 1024 & 9 & 12 & 12\\

HAT-P-11	& 0 & 2012 Jul 27	& \Kpfilt	& 512 & 5 & \nodata & 18\\
& & 2013 Aug 19	& \Kcfilt	& 1024 & 9 & \nodata & 12\\

HAT-P-12	& 0 & 2012 Feb 02	& \Kpfilt	& 1024 & 15 & \nodata & 9\\

HAT-P-13	& 0 & 2012 Feb 02	& \Kpfilt	& 1024 & 9 & \nodata & 9\\

HAT-P-14	& 1 & 2012 Jun 05	& \Kpfilt	& 512 & 5 & 33 & 33\\
& & 2013 Mar 26	& \Hfilt	& 1024 & 9 & 5 & 6\\ 
& & 2013 Mar 26	& \Ksfilt	& 1024 & 10 & 6 & 6\\
& & 2014 Jul 07 & \Ksfilt	& 1024 & 20 & 5 & 5\\

HAT-P-15	& 0 & 2012 Feb 02	& \Kpfilt	& 1024 & 15 & \nodata & 9\\ 
& & 2014 Oct 03	& \Ksfilt	& 1024 & 20 & \nodata & 12\\ 

HAT-P-16	& 1 & 2012 Feb 02	& \Jfilt	& 1024 & 10 & 14 & 15\\
& & 2012 Feb 02	& \Kpfilt	& 1024 & 15 & 15 & 15\\
& & 2013 Aug 19	& \Hfilt	& 1024 & 18 & 6 & 6\\
& & 2013 Aug 19	& \Ksfilt	& 1024 & 9 & 12 & 12\\

HAT-P-17	& 0 & 2012 May 07	& \Kpfilt	& 1024 & 0.9 & \nodata & 18\\ 

HAT-P-18	& 0 & 2012 May 29	& \Kpfilt	& 1024 & 30 & \nodata & 9\\

HAT-P-20	& 0 & 2012 Feb 02	& \Kpfilt	& 512 & 2.5 & \nodata & 18\\
& & 2013 Nov 17	& \Kcfilt	& 1024 & 10 & \nodata & 12\\

HAT-P-22	& 0 & 2012 Feb 02	& \Kpfilt	& 512 & 10 & \nodata & 18\\

HAT-P-24	& 1 & 2012 Feb 02	& \Kpfilt	& 1024 & 15 & 6 & 9\\ 
& & 2014 May 13	& \Jfilt	& 1024 & 10 & 12 & 12\\
& & 2014 May 13	& \Hfilt	& 1024 & 20 & 6 & 6\\
& & 2014 May 13	& \Ksfilt	& 1024 & 10 & 12 & 12\\

HAT-P-26	& 0 & 2012 Feb 02	& \Kpfilt	& 1024 & 15 & \nodata & 9\\

HAT-P-29	& 0 & 2012 Feb 02	& \Jfilt	& 1024 & 10 & \nodata & 9\\
& & 2012 Feb 02	& \Kpfilt	& 1024 & 15 & \nodata & 9\\

HAT-P-30	& 1 & 2012 Feb 02	& \Kpfilt	& 1024 & 10 & 9 & 27\\
& & 2014 May 13	& \Jfilt	& 1024 & 10 & 12 & 12\\ 
& & 2014 May 13	& \Hfilt	& 512 & 10 & 12 & 12\\ 
& & 2014 May 13	& \Ksfilt	& 1024 & 10 & 12 & 12\\ 

HAT-P-31	& 0 & 2012 May 07	& \Kpfilt	& 1024 & 8 & \nodata & 17\\

HAT-P-32	& 1 & 2012 Feb 02	& \Jfilt	& 1024 & 10 & 9 & 9\\
& & 2012 Feb 02	& \Kpfilt	& 1024 & 15 & 15 & 15\\
& & 2013 Mar 02	& \Hfilt	& 1024 & 5 & 9 & 9\\
& & 2013 Mar 02	& \Ksfilt	& 1024 & 15 & 15 & 15\\

HAT-P-33	& 1 & 2012 Feb 02	& \Jfilt	& 1024 & 15 & 8 & 9\\ 
& & 2012 Feb 02	& \Kpfilt	& 1024 & 15 & 9 & 9\\
& & 2013 Mar 02	& \Hfilt	& 1024 & 10 & 12 & 12\\
& & 2013 Mar 02	& \Ksfilt	& 1024 & 10 & 12 & 12\\

HAT-P-34	& 0 & 2012 Jul 27	& \Kpfilt	& 1024 & 9 & \nodata & 30\\

HD-149026 & 0 & 2012 Jun 05 & \Kpfilt	& 256 & 1 & \nodata & 39\\  
& & 2013 Mar 26 & \Ksfilt	& 256 & 9 & \nodata & 12\\ 
& & 2013 Jul 04	& \Hfilt	& 128 & 10 & \nodata & 9\\
& & 2013 Jul 04	& \Ksfilt	& 256 & 10 & \nodata & 12\\
 
TrES-2	& 1 & 2012 Jul 27	& \Jfilt	& 1024 & 9 & 15 & 15\\
& & 2012 Jul 27	& \Hfilt	& 1024 & 9 & 15 & 15\\
& & 2012 Jul 27	& \Kpfilt	& 1024 & 9 & 15 & 15\\
& & 2013 May 31	& \Hfilt	& 1024 & 9 & 12 & 12\\
& & 2013 May 31	& \Ksfilt	& 1024 & 9 & 6 & 6\\

TrES-3	& 0 & 2012 Jul 27	& \Kpfilt	& 1024 & 9 & \nodata & 30\\

TrES-4	& 1 & 2012 Jul 27	& \Jfilt	& 1024 & 9 & 15 & 15\\
& & 2012 Jul 27	& \Hfilt	& 1024 & 9 & 18 & 18\\
& & 2012 Jul 27	& \Kpfilt	& 1024 & 9 & 18 & 18\\
& & 2013 Jul 04	& \Hfilt	& 1024 & 9 & 9 & 9\\
& & 2013 Jul 04	& \Ksfilt	& 1024 & 10 & 12 & 12\\

WASP-1	& 1 & 2012 Jul 27	& \Kpfilt	& 1024 & 20 & 20 & 20\\
& & 2013 Aug 19	& \Hfilt	& 1024 & 9 & 6 & 6\\
& & 2013 Aug 19	& \Ksfilt	& 1024 & 12.5 & 12 & 12\\
& &2014 Jul 12   & \Ksfilt	& 1024 & 15 & 15 & 15\\
& & 2014 Oct 03   & \Ksfilt	& 1024 & 15 & 12 & 12\\

WASP-2	& 1 & 2012 Jul 27	& \Jfilt	& 1024 & 10 & 15 & 15\\
& & 2012 Jul 27	& \Hfilt	& 1024 & 9 & 15 & 15\\
& & 2012 Jul 27	& \Kpfilt	& 1024 & 9 & 30 & 30\\
& & 2013 Jun 22	& \Hfilt	& 1024 & 9 & 11 & 11\\
& & 2013 Jun 22	& \Ksfilt	& 1024 & 9 & 12 & 12\\

WASP-3	& 1 & 2012 Jun 05	& \Kpfilt	& 1024 & 5 & 30 & 30\\
& & 2012 Jul 27	& \Jfilt	& 1024 & 9 & 14 & 15\\
& & 2012 Jul 27	& \Hfilt	& 1024 & 9 & 15 & 15\\
& & 2012 Jul 27	& \Kpfilt	& 1024 & 9 & 30 & 30\\
& & 2013 May 31	& \Hfilt	& 512 & 5 & 5 & 6\\
& & 2013 May 31	& \Ksfilt	& 1024 & 9 & 12 & 12\\

WASP-4	& 0 & 2012 Jul 27	& \Kpfilt	& 1024 & 20 & \nodata & 15\\

WASP-7	& 0 & 2012 Aug 29	& \Kpfilt	& 1024 & 9 & \nodata & 48\\

WASP-8	& 1 & 2012 Jul 27	& \Jfilt	& 1024 & 9 & 10 & 15\\ 
& & 2012 Jul 27	& \Hfilt	& 1024 & 9 & 10 & 15\\ 
& & 2012 Jul 27	& \Kpfilt	& 1024 & 9 & 20 & 30\\ 
& & 2013 Aug 19	& \Ksfilt	& 512 & 12.5 & 8 & 8\\

WASP-10	& 0 & 2012 Jul 04	& \Jfilt	& 1024 & 20 & \nodata & 9\\
& & 2012 Jul 04	& \Kpfilt	& 1024 & 20 & \nodata & 9\\

WASP-12	& 2 & 2012 Feb 02	& \Jfilt	& 1024 & 15 & 9 & 9\\
& & 2012 Feb 02	& \Kpfilt	& 1024 & 15 & 9 & 9\\
& & 2013 Mar 02	& \Ksfilt	& 1024 & 10 & 15 & 15\\

WASP-14	& 1 & 2012 Jul 27	& \Jfilt	& 1024 & 9 & 15 & 15\\
& & 2012 Jul 27	& \Hfilt	& 1024 & 9 & 15 & 15\\
& & 2012 Jul 27	& \Kpfilt	& 1024 & 9 & 33 & 33\\
& & 2013 Mar 26	& \Kpfilt	& 1024 & 9 & 12 & 12\\
& & 2013 Mar 26	& \Ksfilt	& 1024 & 9 & 12 & 12\\

WASP-15	& 0 & 2012 Feb 02	& \Kpfilt	& 1024 & 15 & \nodata & 9\\

WASP-16	& 0 & 2012 Feb 02	& \Kpfilt	& 1024 & 15 & \nodata & 18\\
& & 2012 Jul 27	& \Hfilt	& 1024 & 9 & \nodata & 15\\ 
& & 2012 Jul 27	& \Kpfilt	& 1024 & 9 & \nodata & 33\\

WASP-17	& 0 & 2012 May 07	& \Kpfilt	& 1024 & 10 & \nodata & 18\\

WASP-18	& 0 & 2012 Aug 28	& \Kpfilt	& 1024 & 9 & \nodata & 32\\ 

WASP-22	& 0 & 2012 Aug 26	& \Kpfilt	& 1024 & 10 & \nodata & 9\\

WASP-24	& 0 & 2012 Feb 02	& \Kpfilt	& 1024 & 15 & \nodata & 9\\
& & 2012 Jun 03	& \Kpfilt	& 1024 & 15 & \nodata & 12\\

WASP-34	& 0 & 2012 Feb 02	& \Kpfilt	& 1024 & 10 & \nodata & 18\\

WASP-38	& 0 & 2012 Jun 04	& \Kpfilt	& 1024 & 30 & \nodata & 4\\

XO-2		& 0 & 2012 Feb 02	& \Kpfilt	& 1024 & 10 & \nodata & 27\\

XO-3		& 0 & 2012 Feb 02	& \Jfilt	& 512 & 2.5 & \nodata & 9\\  
& & 2012 Feb 02	& \Kpfilt	& 512 & 2.5 & \nodata & 9\\
& & 2013 Mar 02	& \Hfilt	& 1024 & 15 & \nodata & 12\\
& & 2013 Mar 02	& \Kpfilt	& 1024 & 15 & \nodata & 9\\

XO-4		& 0 & 2012 Feb 02	& \Kpfilt	& 1024 & 9 & \nodata & 9\\

XO-5		& 0 & 2012 Feb 02	& \Kpfilt	& 1024 & 15 & \nodata & 16
\enddata
\tablecomments{The $N_c$ columns indicates the number of confirmed common proper motion companions found for each target. The ``Array'' column denotes the horizontal width, in pixels, of the section of the NIRC2 detector used to capture the image. The array dimensions used in this survey were 1024x1024 (the full NIRC2 array), 512x512, 256x264, or 128x152. These dimensions are constrained by NIRC2's readout software. The $T_{\mathrm{int}}$ column indicates the total integration time of a single exposure, in seconds. For targets with detected companions, the $N_{\mathrm{fit}}$ column indicates the number of exposures used to make our photometric and/or astrometric measurements of companions. The $N_{\mathrm{stack}}$ column indicates the number of exposures used to compute the final stacked image for contrast curve measurements. The last column is only given for images taken in the \Kpfilt, \Ksfilt, or \Kcfilt bandpasses because we only compute contrast curves in these bands. $N_{\mathrm{fit}}$ and $N_{\mathrm{stack}}$ may differ because the companion may not be visible at all dither positions and/or poor conditions prevented acquisition of useful data.}
\end{deluxetable}

\clearpage
\begin{deluxetable*}{lccccccccc}
\tablecolumns{10}
\tabletypesize{\scriptsize}
\tablecaption{Target stellar parameters\label{tab:stellar_params}}
\tablehead{
& & & \colhead{$T_{\mathrm{eff}}$}  & \colhead{$M$} & \colhead{$\log g$} & \colhead{$D$}& \multicolumn{3}{c}{References for...} \vspace{-0.1cm} \\
\colhead{Target} & \colhead{$N_c$} & \colhead{Sample} & \vspace{-0.2cm}& & & & & & \\
 & & & \colhead{(K)} & \colhead{(\Msun)} & \colhead{(cgs)} & \colhead{(pc)} & \colhead{$T$} & \colhead{$M,\log g$} & \colhead{$D$}
}
\startdata
GJ 436     & 0 & Misaligned & $3416 \pm 54$ & $0.507 \pm 0.070$  & $4.83 \pm 0.03$             & $10.14 \pm 0.24$ & 1 & 1 & 2\\
HAT-P-2   & 0 & Misaligned & $6290 \pm 60$ & $1.36 \pm 0.04$     & $4.138 \pm 0.035$        & $125.3 \pm 13.1$ & 3 & 3 & 2\\
HAT-P-4   & 0 & Control       & $5890 \pm 67$ & $1.26 \pm 0.10$       & $4.12 \pm 0.03$            & $293.5 \pm 19.4$ & 4 & 5,6 & 2\\
HAT-P-6   & 0 & Misaligned & $6687 \pm 71$ & $1.29 \pm 0.06$     & $4.188 \pm 0.035$        & $277.8 \pm 19.1$ & 4 & 7 & 2 \\
HAT-P-7   & 1 & Misaligned & $6259 \pm 32$ & $1.361 \pm 0.021$ & $4.98\pm0.13$              & $320^{+50}_{-40}$ & 8 & 8 & 2 \\
HAT-P-8   & 2 & Control      & $6223 \pm 67$ & $1.192 \pm 0.075$ & $4.177 \pm 0.022$         & $230 \pm 15$ & 4 & 9 & 2 \\
HAT-P-10 & 1 & Control      & $4974 \pm 65$ & $0.83 \pm 0.03$     & $4.56 \pm 0.02$             & $121.7 \pm 4.2$ & 4 & 10 & 10 \\
HAT-P-11 & 0 & Misaligned & $4792 \pm 69$ & $0.81 \pm 0.03$     & $4.59 \pm 0.03$             & $38.0 \pm 1.3$ & 4 & 11 & 11\\
HAT-P-12 & 0 & Control      & $4650 \pm 60$ & $0.733 \pm 0.018$ & $4.61 \pm 0.01$             & $139.1 \pm 9.6$ & 12 & 12 & 2 \\
HAT-P-13 & 0 & Misaligned & $5720 \pm 69$ & $1.320 \pm 0.062$ & $4.070 \pm 0.020$        & $214 \pm 12$ & 4 & 13 & 14 \\
HAT-P-14 & 1 & Misaligned & $6671 \pm 75$ & $1.418 \pm 0.054$ & $4.187\pm0.025$          & $205 \pm 11$ & 4 & 7 & 15\\
HAT-P-15 & 0 & Misaligned & $5640 \pm 69$ & $1.013 \pm 0.043$ & $4.38 \pm 0.03$            & $190 \pm 8$ & 4 & 16 & 16\\ 
HAT-P-16 & 1 & Misaligned & $6140 \pm 72$ & $1.218 \pm 0.039$ & $4.34 \pm 0.03$            & $235 \pm 10$ & 4 & 17 & 17 \\
HAT-P-17 & 0 & Misaligned & $5345 \pm 70$ & $0.857 \pm 0.039$ & $4.52 \pm 0.02$            & $90 \pm 3$ & 4 & 18 & 18 \\
HAT-P-18 & 0 & Control      & $4790 \pm 72$ & $0.770 \pm 0.031$ & $4.57 \pm 0.04$             & $166 \pm 9$ & 4 & 19 & 19 \\
HAT-P-20 & 0 & Misaligned & $4619 \pm 72$ & $0.756 \pm 0.028$ & $4.63 \pm 0.02$            & $70 \pm 3$ & 4 & 20 & 21 \\
HAT-P-22 & 0 & Control      & $5367 \pm 67$ & $0.916 \pm 0.035$ & $4.36 \pm 0.04$             & $82 \pm 3$ & 4 & 20 & 21 \\
HAT-P-24 & 1 & Control      & $6329 \pm 67$ & $1.191 \pm 0.042$ & $4.27 \pm 0.04$             & $396 \pm 20$ & 4 & 22 & 22\\
HAT-P-26 & 0 & Control      & $5142 \pm 69$ & $0.816 \pm 0.033$ & $4.56 \pm 0.06$             & $134^{+18}_{-8}$ & 4 & 23 & 23 \\
HAT-P-29 & 0 & Control      & $6086 \pm 69$ & $1.207 \pm 0.046$ & $4.34 \pm 0.06$             & $322^{+35}_{-21}$ & 4 & 24 & 24 \\ 
HAT-P-30 & 1 & Misaligned & $6304 \pm 88$ & $1.242 \pm 0.041$ & $4.36 \pm 0.03$            & $193 \pm 8$ & 25 & 25 & 25\\
HAT-P-31 & 0 & Misaligned & $6065 \pm 100$ & $1.218 \pm 0.070$ & $4.26^{+0.11}_{-0.13}$ &$354^{+74}_{-51}$ & 26 & 26 & 26\\
HAT-P-32 & 1 & Misaligned & $6207 \pm 88$ & $1.160 \pm 0.041$ & $4.33 \pm 0.01$            & $283 \pm 5$ & 27 & 27 & 27\\
HAT-P-33 & 1 & Control      & $6446 \pm 88$ & $1.375 \pm 0.040$ & $4.15 \pm 0.01$             & $387 \pm 9$ & 27 & 27 & 27\\
HAT-P-34 & 0 & Misaligned & $6442 \pm 88$ & $1.392 \pm 0.047$ & $4.21\pm0.06$              & $257^{+22}_{-17}$ & 28 & 28 & 28\\
HD149026 & 0 & Control    & $6103 \pm 66$ & $1.345 \pm 0.020$  & $4.189\pm0.020$          &  $80.8 \pm 4.0$ & 4 & 29 & 2\\
TrES-2     & 1 & Control      & $5850 \pm 50$ & $0.94 \pm 0.05$     & $4.45 \pm 0.01$             &  $220 \pm 10\tablenotemark{a}$ & 30 & 31 & 2\\
TrES-3     & 0 & Misaligned & $5514 \pm 69$ & $0.928 \pm 0.038$ & $4.57 \pm 0.01$            &  $258.5 \pm 16.1$\tablenotemark{a} & 4 & 32 & 2\\
TrES-4     & 1 & Control      & $6200 \pm 75$ & $1.339\pm0.086$ & $4.030\pm0.033$             & $476 \pm 26\tablenotemark{a}$ & 32 & 7 & 2\\
WASP-1   & 1 & Control      & $6160 \pm 64$ & $1.265 \pm 0.054$    & $4.209 \pm 0.051$      & $380 \pm 38\tablenotemark{a}$ & 4 & 7 & 2\\
WASP-2   & 1 & Misaligned & $5255 \pm 71$ & $0.851 \pm 0.050$  & $4.537 \pm 0.017$         &  $140 \pm 10$\tablenotemark{a} & 4 & 7 & 2\\
WASP-3   & 1 & Control      & $6375 \pm 63$ & $1.20 \pm 0.01$    & $4.33\pm0.03$                &  $220 \pm 20\tablenotemark{a}$ & 4 & 33 & 2\\
WASP-4   & 0 & Control      & $5540 \pm 55$ & $0.927 \pm 0.056$    & $4.487 \pm 0.010$     &  $280.9 \pm 31.1$ & 34 & 7 & 2\\
WASP-7   & 0 & Misaligned & $6520 \pm 70$ & $1.317 \pm 0.072$   & $4.218 \pm 0.048$      & $140 \pm 15$ & 34 & 7 & 35\\
WASP-8   & 1 & Misaligned & $5570 \pm 85$ & $1.04 \pm 0.08$    & $4.41 \pm 0.09$             & $87 \pm 7$ & 34 & 36 & 2\\
WASP-10 & 0 & Misaligned & $4735 \pm 69$ & $0.703 \pm 0.070$ & $4.51^{+0.06}_{-0.05}$  & $90 \pm 20$ & 4 & 37 & 37\\
WASP-12 & 2 & Misaligned & $6118 \pm 64$ & $1.38 \pm 0.19$      & $4.159 \pm 0.024$       & $427 \pm 90\tablenotemark{a}$ & 4 & 7 & 2\\
WASP-14 & 1 & Misaligned & $6462 \pm 75$ & $1.350 \pm 0.121$    & $4.126 \pm 0.042$     & $160 \pm 20\tablenotemark{a}$ & 4 & 7 & 2\\
WASP-15 & 0 & Misaligned & $6405 \pm 80$ & $1.305 \pm 0.051$ & $4.189 \pm 0.021$        & $256 \pm 32\tablenotemark{b}$ & 36 & 38 & See note\tablenotemark{b}\\
WASP-16 & 0 & Control       & $5630 \pm 70$ & $0.980 \pm 0.054$ & $4.357 \pm 0.022$        & $174 \pm 14\tablenotemark{b}$ & 36 & 38 & See note\tablenotemark{b}\\
WASP-17 & 0 & Misaligned & $6550 \pm 100$ & $1.286 \pm 0.079$ & $4.149 \pm 0.014$      & $476 \pm 36\tablenotemark{a}$ & 39 & 39 & 2\\
WASP-18 & 0 & Control       & $6368 \pm 66$ & $1.274 \pm 0.060$     & $4.365 \pm 0.022$   & $122.6 \pm 6.7\tablenotemark{a}$ & 4 & 7 & 2\\
WASP-22 & 0 & Control       & $5958 \pm 98$ & $1.109 \pm 0.026$ & $4.31 \pm 0.03$           & $300 \pm 30\tablenotemark{c}$ & 40 & 40 & 41\tablenotemark{c}\\
WASP-24 & 0 & Control       & $6107 \pm 77$ & $1.184 \pm 0.027$ & $4.263 \pm 0.017$       & $332.5 \pm 23.8\tablenotemark{a}$ & 4 & 42 & 2\\
WASP-34 & 0 & Control       & $5700 \pm 100$ & $1.01 \pm 0.07$   & $4.5 \pm 0.1$               & $120 \pm 15$ & 43 & 43 & 43\\
WASP-38 & 0 & Misaligned & $6187 \pm 77$ & $1.23 \pm 0.04$     & $4.267 \pm 0.030$         & $110 \pm 20$ & 4 & 44 & 45\\
XO-2        & 0 & Control       & $5377 \pm 79$ & $0.924 \pm 0.173$     & $4.436 \pm 0.042$   & $156.0 \pm 8.8$ & 4 & 7 & 2\\
XO-3        & 0 & Misaligned & $6759 \pm 79$ & $1.213 \pm 0.066$ & $4.244 \pm 0.041$       & $185.7 \pm 11.8$ & 4 & 46 & 2\\
XO-4        & 0 & Misaligned & $6297 \pm 72$ & $1.32 \pm 0.02$     & $4.18 \pm 0.07$           & $308.2 \pm 19.6$ & 4 & 47 & 2\\
XO-5        & 0 & Control       & $5370 \pm 70$ & $0.88 \pm 0.03$    & $4.31 \pm 0.03$            &  $260 \pm 12$ & 48 & 48 & 48
\enddata
\vspace{-0.1cm}
\tablecomments{$N_c$ is the number of confirmed common proper motion companions found for each target. ``Sample'' shows whether we placed the target in the ``misaligned'' (planets on misaligned or eccentric orbits) or the ``control'' (planets on well-aligned and circular orbits) subsamples. With the exception of GJ 436 and HAT-P-2, our stars do not have directly measured parallax estimates.  We take our distance estimates from the referenced papers, which combine estimates of the stellar surface gravity, effective temperature, and metallicity from high-resolution optical spectroscopy and (in some cases) constraints on the stellar density from fits to the transit light curve in order to constrain the mass, radius, and age of the host star. The measured apparent magnitudes in $V$, $J$, $H$, and $K$ bands can then be used to estimate the distance to the star given these known properties. The final three columns lists references for our temperature; mass and $\log g$; and distance measurements, respectively.}
\tablerefs{(1) \citet{vonBraun2012}; (2) \citet{Triaud2014}; (3) \citet{Pal2010}; (4) \citet{Torres2012}; (5) \citet{Kovacs2007}; (6) \citet{Winn2011}; (7) \citet{Southworth2012b}; (8) \citet{vanEylen2012}; (9) \citet{Mancini2013}; (10) \citet{Bakos2009a}; (11) \citet{Bakos2010}; (12) \citet{Hartman2009}; (13) \citet{Southworth2012a}; (14) \citet{Bakos2009b}; (15) \citet{Torres2010}; (16) \citet{Kovacs2010}; (17) \citet{Buchhave2010}; (18) \citet{Howard2012a}; (19) \citet{Hartman2011a}; (20) \citet{Bakos2011}; (21) \citet{Bakos2010}; (22) \citet{Kipping2010}; (23) \citet{Hartman2011b}; (24) \citet{Buchhave2011}; (25) \citet{Johnson2011}; (26) \citet{Kipping2011}; (27) \citet{Hartman2011c}; (28) \citet{Bakos2012}; (29) \citet{Carter2009}; (30) \citet{Sozzetti2007}; (31) \citet{Barclay2012}; (32) \citet{Sozzetti2009}; (33) \citet{Miller2010}; (34) \citet{Maxted2011}; (35) \citet{Hellier2009a}; (36) \citet{Doyle2013}; (37) \citet{Christian2009}; (38) \citet{Southworth2013}; (39) \citet{Southworth2012c}; (40) \citet{Anderson2011}; (41) \citet{Maxted2010}; (42) \citet{Street2010}; (43) \citet{Smalley2011}; (44) \citet{Brown2012}; (45) \citet{Barros2011}; (46) \citet{Winn2008b}; (47) \citet{McCullough2008}; (48) \citet{Pal2009}}
\tablenotetext{a}{This distance estimate has been updated from the values used in the first ``Friends of Hot Jupiters'' paper~\citep{Knutson2014}. The new distances are consistent with the previously used values except for WASP-12.}
\tablenotetext{b}{We could not find distance measurements for these targets so we estimated the distance based on relative brightness of the target to a reference star and used the same relative error. For WASP-15, we used WASP-14 as a reference and for WASP-16, we used WASP-8 as a reference.}
\tablenotetext{c}{A distance was reported with no uncertainty, so we use a conservative estimate of 10\%, based on uncertainties of other stars.}
\end{deluxetable*}

\vspace{-0.8cm}
\begin{deluxetable}{llcccc}
\tablecolumns{6}
\tablecaption{Flux ratio measurements of confirmed common proper motion companions  \label{tab:comp_phot}}
\tablehead{
\colhead{Companion} & \colhead{UT Obs. Date} & \colhead{$\Delta \Jfilt$} & \colhead{$\Delta \Hfilt$} & \colhead{$\Delta \Kpfilt$} & \colhead{$\Delta \Ksfilt$}
}
\startdata
HAT-P-7B  & 2012 Jul 27 & $6.562 \pm 0.075$ & $5.887 \pm 0.078$ & $5.920 \pm 0.058$ & \nodata \\
HAT-P-7B  & 2013 May 31 & \nodata & \nodata & \nodata & $5.948 \pm 0.040$ \\
HAT-P-7B  & 2013 Jun 22 & \nodata & $5.926 \pm 0.093$ & \nodata & $6.04 \pm 0.12$ \\
HAT-P-7B  & 2014 Jul 12 & \nodata & \nodata & \nodata & $5.827 \pm 0.058$ \\
HAT-P-8B  & 2012 Jul 27 & $6.59 \pm 0.12$ & $5.671 \pm 0.054$ & $5.874 \pm 0.063$ & \nodata \\
HAT-P-8C  & 2012 Jul 27 & $7.16 \pm 0.15$ & $6.352 \pm 0.065$ & $6.447 \pm 0.057$ & \nodata \\
HAT-P-8B  & 2013 Aug 19 & \nodata & $5.597 \pm 0.036$ & $5.841 \pm 0.073$ & \nodata \\
HAT-P-8C  & 2013 Aug 19 & \nodata & $6.147 \pm 0.077$ & $6.310 \pm 0.057$ & \nodata \\
HAT-P-8B  & 2013 Aug 19 & \nodata & $5.597 \pm 0.036$ & \nodata & $5.726 \pm 0.075$ \\
HAT-P-8C  & 2013 Aug 19 & \nodata & $6.147 \pm 0.077$ & \nodata & $6.201 \pm 0.094$ \\
HAT-P-10B & 2012 Feb 02 & $2.656 \pm 0.017$ & \nodata & $2.960 \pm 0.022$ & \nodata \\
HAT-P-10B & 2013 Aug 19 & \nodata & $2.448 \pm 0.014$ & $2.763 \pm 0.034$ & \nodata \\
HAT-P-10B & 2013 Aug 19 & \nodata & $2.448 \pm 0.014$ & \nodata & $2.733 \pm 0.045$ \\
HAT-P-14B & 2012 Jun 05 & \nodata & \nodata & $5.633 \pm 0.033$ & \nodata \\
HAT-P-14B & 2013 Mar 26 & \nodata & $5.237 \pm 0.086$ & \nodata & $5.647 \pm 0.096$ \\
HAT-P-14B & 2014 Jul 07 & \nodata & \nodata & \nodata & $5.844 \pm 0.034$ \\
HAT-P-16B & 2012 Feb 02 & $5.421 \pm 0.039$ & \nodata & $5.530 \pm 0.021$ & \nodata \\
HAT-P-16B & 2013 Aug 19 & \nodata & $5.147 \pm 0.047$ & \nodata & $5.382 \pm 0.061$ \\
HAT-P-24B & 2012 Feb 02 & \nodata & \nodata & $4.45 \pm 0.12$ & \nodata \\
HAT-P-24B & 2014 May 13 & $4.658 \pm 0.080$ & $4.163 \pm 0.052$ & \nodata & $3.811 \pm 0.010$ \\
HAT-P-30B & 2012 Feb 02 & \nodata & \nodata & $3.134 \pm 0.054$ & \nodata \\
HAT-P-30B & 2014 May 13 & $3.4304 \pm 0.0063$ & $3.1153 \pm 0.0053$ & \nodata & $2.975 \pm 0.041$ \\
HAT-P-32B & 2012 Feb 02 & $4.148 \pm 0.028$ & \nodata & $3.733 \pm 0.042$ & \nodata \\
HAT-P-32B & 2013 Mar 02 & \nodata & $3.668 \pm 0.071$ & \nodata & $3.500 \pm 0.073$ \\
HAT-P-33B & 2012 Feb 02 & $4.137 \pm 0.054$ & \nodata & $3.938 \pm 0.085$ & \nodata \\
HAT-P-33B & 2013 Mar 02 & \nodata & $3.469 \pm 0.033$ & \nodata & $3.415 \pm 0.057$ \\
TrES-2B   & 2012 Jul 27 & $2.896 \pm 0.020$ & $2.4515 \pm 0.0073$ & $2.422 \pm 0.019$ & \nodata \\
TrES-2B   & 2013 May 31 & \nodata & $2.4628 \pm 0.0085$ & \nodata & $2.523 \pm 0.036$ \\
TrES-4B   & 2012 Jul 27 & $3.528 \pm 0.012$ & $3.177 \pm 0.013$ & $3.301 \pm 0.046$ & \nodata \\
TrES-4B   & 2013 Jul 04 & \nodata & $3.1841 \pm 0.0084$ & \nodata & $3.447 \pm 0.024$ \\
WASP-1B   & 2012 Jul 27 & \nodata & \nodata & $4.551 \pm 0.077$ & \nodata \\
WASP-1B   & 2013 Aug 19 & \nodata & $4.766 \pm 0.046$ & \nodata & $4.840 \pm 0.055$ \\
WASP-1B   & 2014 Jul 12 & \nodata & \nodata & \nodata & $4.858 \pm 0.064$ \\
WASP-1B   & 2014 Oct 03 & \nodata & \nodata & \nodata & $4.561 \pm 0.100$ \\
WASP-2B   & 2012 Jul 27 & $2.880 \pm 0.010$ & $2.7343 \pm 0.0061$ & $2.871 \pm 0.028$ & \nodata \\
WASP-2B   & 2013 Jun 22 & \nodata & $2.7254 \pm 0.0075$ & \nodata & $2.687 \pm 0.017$ \\
WASP-3B   & 2012 Jun 05 & \nodata & \nodata & $6.528 \pm 0.049$ & \nodata \\
WASP-3B   & 2012 Jul 27 & $7.27 \pm 0.12$ & $6.683 \pm 0.079$ & $6.641 \pm 0.052$ & \nodata \\
WASP-3B   & 2013 May 31 & \nodata & $6.550 \pm 0.091$ & \nodata & $6.552 \pm 0.027$ \\
WASP-8B   & 2012 Jul 27 & $4.492 \pm 0.062$ & $3.134 \pm 0.056$ & $2.550 \pm 0.021$ & \nodata \\
WASP-8B   & 2013 Aug 19 & \nodata & \nodata & \nodata & $2.560 \pm 0.018$ \\
WASP-12B  & 2012 Feb 02 & $3.711 \pm 0.076$ & \nodata & $3.32 \pm 0.11$ & \nodata \\
WASP-12C  & 2012 Feb 02 & $3.672 \pm 0.021$ & \nodata & $3.577 \pm 0.036$ & \nodata \\
WASP-12B  & 2013 Mar 02 & \nodata & \nodata & \nodata & $3.286 \pm 0.030$ \\
WASP-12C  & 2013 Mar 02 & \nodata & \nodata & \nodata & $3.182 \pm 0.017$ \\
WASP-14B  & 2012 Jul 27 & $5.207 \pm 0.020$ & $4.788 \pm 0.024$ & $4.984 \pm 0.025$ & \nodata \\
WASP-14B  & 2013 Mar 26 & \nodata & \nodata & $4.765 \pm 0.057$ & \nodata \\
WASP-14B  & 2013 Mar 26 & \nodata & \nodata & \nodata & $4.701 \pm 0.054$
\enddata
\tablecomments{$\Delta X$ is the difference in magnitude between the companion and primary stars in the X filter.}
\end{deluxetable}

\clearpage
\begin{deluxetable}{llccccccc}
\tablecolumns{9}
\tablecaption{Multi-band photometry of confirmed common proper motion companions \label{tab:phot_colors}}
\tablehead{
\colhead{Companion} & \colhead{UT Obs. Date} & \colhead{K} & \colhead{$m_J$} & \colhead{$m_H$} & \colhead{$m_K$} & \colhead{$J-K$} & \colhead{$H-K$} & \colhead{$J-H$} 
}
\startdata
HAT-P-7B  & 2012 Jul 27 & \Kpfilt & $16.117 \pm 0.075$ & $15.231 \pm 0.078$ & $15.254 \pm 0.058$ & $0.863 \pm 0.095$ & $-0.023 \pm 0.097$ & $0.89 \pm 0.11$ \\
HAT-P-7B  & 2013 May 31 & \Ksfilt & \nodata & \nodata & $15.282 \pm 0.040$ & \nodata & \nodata & \nodata \\
HAT-P-7B  & 2013 Jun 22 & \Ksfilt & \nodata & $15.270 \pm 0.093$ & $15.37 \pm 0.12$ & \nodata & $-0.10 \pm 0.15$ & \nodata \\
HAT-P-7B  & 2014 Jul 12 & \Ksfilt & \nodata & \nodata & $15.161 \pm 0.058$ & \nodata & \nodata & \nodata \\
HAT-P-8B  & 2012 Jul 27 & \Kpfilt & $15.81 \pm 0.12$ & $14.675 \pm 0.054$ & $14.827 \pm 0.063$ & $0.98 \pm 0.14$ & $-0.152 \pm 0.083$ & $1.13 \pm 0.13$ \\
HAT-P-8C  & 2012 Jul 27 & \Kpfilt & $16.37 \pm 0.15$ & $15.356 \pm 0.065$ & $15.400 \pm 0.057$ & $0.97 \pm 0.16$ & $-0.044 \pm 0.087$ & $1.02 \pm 0.16$ \\
HAT-P-8B  & 2013 Aug 19 & \Kpfilt & \nodata & $14.601 \pm 0.036$ & $14.794 \pm 0.073$ & \nodata & $-0.193 \pm 0.082$ & \nodata \\
HAT-P-8C  & 2013 Aug 19 & \Kpfilt & \nodata & $15.151 \pm 0.077$ & $15.263 \pm 0.057$ & \nodata & $-0.113 \pm 0.096$ & \nodata \\
HAT-P-8B  & 2013 Aug 19 & \Ksfilt & \nodata & $14.601 \pm 0.036$ & $14.679 \pm 0.075$ & \nodata & $-0.078 \pm 0.083$ & \nodata \\
HAT-P-8C  & 2013 Aug 19 & \Ksfilt & \nodata & $15.151 \pm 0.077$ & $15.154 \pm 0.094$ & \nodata & $-0.00 \pm 0.12$ & \nodata \\
HAT-P-10B & 2012 Feb 02 & \Kpfilt & $12.671 \pm 0.017$ & \nodata & $12.381 \pm 0.022$ & $0.290 \pm 0.027$ & \nodata & \nodata \\
HAT-P-10B & 2013 Aug 19 & \Kpfilt & \nodata & $12.008 \pm 0.014$ & $12.184 \pm 0.034$ & \nodata & $-0.176 \pm 0.037$ & \nodata \\
HAT-P-10B & 2013 Aug 19 & \Ksfilt & \nodata & $12.008 \pm 0.014$ & $12.154 \pm 0.045$ & \nodata & $-0.146 \pm 0.047$ & \nodata \\
HAT-P-14B & 2012 Jun 05 & \Kpfilt & \nodata & \nodata & $14.484 \pm 0.033$ & \nodata & \nodata & \nodata \\
HAT-P-14B & 2013 Mar 26 & \Ksfilt & \nodata & $14.164 \pm 0.086$ & $14.498 \pm 0.096$ & \nodata & $-0.33 \pm 0.13$ & \nodata \\
HAT-P-14B & 2014 Jul 07 & \Ksfilt & \nodata & \nodata & $14.695 \pm 0.034$ & \nodata & \nodata & \nodata \\
HAT-P-16B & 2012 Feb 02 & \Kpfilt & $15.271 \pm 0.039$ & \nodata & $15.083 \pm 0.021$ & $0.189 \pm 0.045$ & \nodata & \nodata \\
HAT-P-16B & 2013 Aug 19 & \Ksfilt & \nodata & $14.770 \pm 0.047$ & $14.935 \pm 0.061$ & \nodata & $-0.165 \pm 0.077$ & \nodata \\
HAT-P-24B & 2012 Feb 02 & \Kpfilt & \nodata & \nodata & $14.99 \pm 0.12$ & \nodata & \nodata & \nodata \\
HAT-P-24B & 2014 May 13 & \Ksfilt & $15.455 \pm 0.080$ & $14.752 \pm 0.052$ & $14.354 \pm 0.010$ & $1.101 \pm 0.081$ & $0.398 \pm 0.053$ & $0.703 \pm 0.096$ \\
HAT-P-30B & 2012 Feb 02 & \Kpfilt & \nodata & \nodata & $12.285 \pm 0.054$ & \nodata & \nodata & \nodata \\
HAT-P-30B & 2014 May 13 & \Ksfilt & $12.8724 \pm 0.0063$ & $12.3353 \pm 0.0053$ & $12.126 \pm 0.041$ & $0.746 \pm 0.042$ & $0.209 \pm 0.042$ & $0.5371 \pm 0.0082$ \\
HAT-P-32B & 2012 Feb 02 & \Kpfilt & $14.399 \pm 0.028$ & \nodata & $13.723 \pm 0.042$ & $0.676 \pm 0.050$ & \nodata & \nodata \\
HAT-P-32B & 2013 Mar 02 & \Ksfilt & \nodata & $13.692 \pm 0.071$ & $13.490 \pm 0.073$ & \nodata & $0.20 \pm 0.10$ & \nodata \\
HAT-P-33B & 2012 Feb 02 & \Kpfilt & $14.400 \pm 0.054$ & \nodata & $13.942 \pm 0.085$ & $0.46 \pm 0.10$ & \nodata & \nodata \\
HAT-P-33B & 2013 Mar 02 & \Ksfilt & \nodata & $13.530 \pm 0.033$ & $13.419 \pm 0.057$ & \nodata & $0.111 \pm 0.066$ & \nodata \\
TrES-2B   & 2012 Jul 27 & \Kpfilt & $13.128 \pm 0.020$ & $12.3715 \pm 0.0073$ & $12.268 \pm 0.019$ & $0.859 \pm 0.028$ & $0.103 \pm 0.021$ & $0.756 \pm 0.021$ \\
TrES-2B   & 2013 May 31 & \Ksfilt & \nodata & $12.3828 \pm 0.0085$ & $12.369 \pm 0.036$ & \nodata & $0.014 \pm 0.037$ & \nodata \\
TrES-4B   & 2012 Jul 27 & \Kpfilt & $14.111 \pm 0.012$ & $13.527 \pm 0.013$ & $13.631 \pm 0.046$ & $0.481 \pm 0.048$ & $-0.104 \pm 0.048$ & $0.584 \pm 0.018$ \\
TrES-4B   & 2013 Jul 04 & \Ksfilt & \nodata & $13.5341 \pm 0.0084$ & $13.777 \pm 0.024$ & \nodata & $-0.243 \pm 0.025$ & \nodata \\
WASP-1B   & 2012 Jul 27 & \Kpfilt & \nodata & \nodata & $14.827 \pm 0.077$ & \nodata & \nodata & \nodata \\
WASP-1B   & 2013 Aug 19 & \Ksfilt & \nodata & $15.130 \pm 0.046$ & $15.116 \pm 0.055$ & \nodata & $0.014 \pm 0.072$ & \nodata \\
WASP-1B   & 2014 Jul 12 & \Ksfilt & \nodata & \nodata & $15.134 \pm 0.064$ & \nodata & \nodata & \nodata \\
WASP-1B   & 2014 Oct 03 & \Ksfilt & \nodata & \nodata & $14.837 \pm 0.100$ & \nodata & \nodata & \nodata \\
WASP-2B   & 2012 Jul 27 & \Kpfilt & $13.046 \pm 0.010$ & $12.4863 \pm 0.0061$ & $12.503 \pm 0.028$ & $0.543 \pm 0.030$ & $-0.016 \pm 0.028$ & $0.560 \pm 0.012$ \\
WASP-2B   & 2013 Jun 22 & \Ksfilt & \nodata & $12.4774 \pm 0.0075$ & $12.319 \pm 0.017$ & \nodata & $0.158 \pm 0.018$ & \nodata \\
WASP-3B   & 2012 Jun 05 & \Kpfilt & \nodata & \nodata & $15.889 \pm 0.049$ & \nodata & \nodata & \nodata \\
WASP-3B   & 2012 Jul 27 & \Kpfilt & $16.88 \pm 0.12$ & $16.090 \pm 0.079$ & $16.002 \pm 0.052$ & $0.87 \pm 0.13$ & $0.088 \pm 0.095$ & $0.78 \pm 0.15$ \\
WASP-3B   & 2013 May 31 & \Ksfilt & \nodata & $15.957 \pm 0.091$ & $15.913 \pm 0.027$ & \nodata & $0.044 \pm 0.095$ & \nodata \\
WASP-8B   & 2012 Jul 27 & \Kpfilt & $12.993 \pm 0.062$ & $11.354 \pm 0.056$ & $10.636 \pm 0.021$ & $2.357 \pm 0.065$ & $0.719 \pm 0.060$ & $1.638 \pm 0.083$ \\
WASP-8B   & 2013 Aug 19 & \Ksfilt & \nodata & \nodata & $10.646 \pm 0.018$ & \nodata & \nodata & \nodata \\
WASP-12B  & 2012 Feb 02 & \Kpfilt & $14.188 \pm 0.076$ & \nodata & $13.51 \pm 0.11$ & $0.68 \pm 0.13$ & \nodata & \nodata \\
WASP-12C  & 2012 Feb 02 & \Kpfilt & $14.149 \pm 0.021$ & \nodata & $13.765 \pm 0.036$ & $0.384 \pm 0.042$ & \nodata & \nodata \\
WASP-12B  & 2013 Mar 02 & \Ksfilt & \nodata & \nodata & $13.474 \pm 0.030$ & \nodata & \nodata & \nodata \\
WASP-12C  & 2013 Mar 02 & \Ksfilt & \nodata & \nodata & $13.370 \pm 0.017$ & \nodata & \nodata & \nodata \\
WASP-14B  & 2012 Jul 27 & \Kpfilt & $14.076 \pm 0.020$ & $13.438 \pm 0.024$ & $13.605 \pm 0.025$ & $0.471 \pm 0.032$ & $-0.167 \pm 0.034$ & $0.638 \pm 0.031$ \\
WASP-14B  & 2013 Mar 26 & \Kpfilt & \nodata & \nodata & $13.386 \pm 0.057$ & \nodata & \nodata & \nodata \\
WASP-14B  & 2013 Mar 26 & \Ksfilt & \nodata & \nodata & $13.322 \pm 0.054$ & \nodata & \nodata & \nodata
\enddata
\tablecomments{The K column indicates whether the following columns refer to the \Kpfilt\ or \Ksfilt\ bandpass, as both were used in the Friends of Hot Jupiters survey. The $m_X$ columns refer to the secondary star's apparent magnitude in the X filter, which is computed from the primary star's apparent magnitude and the measured flux ratio as reported in Table~\ref{tab:comp_phot}. The last three columns show the computed color of the companion star.}
\tablerefs{Primary star apparent magnitudes from 2MASS~\citep{Skrutskie2006}.}
\end{deluxetable}

\clearpage
\begin{deluxetable}{llccc}
\tablecolumns{5}
\tablecaption{K-band astrometric measurements of confirmed common proper motion companions \label{tab:comp_astr}}
\tablehead{
& & & \colhead{$\rho$} & \colhead{PA} \vspace{-0.1cm}\\
\colhead{Companion} & \colhead{UT Obs. Date} & \colhead{K} &\vspace{-0.2cm}  & \\
& & & \colhead{(mas)} & \colhead{(\deg)}
}
\startdata
HAT-P-7B  & 2012 Jul 27 & \Kpfilt & $3858.7 \pm 1.8$ & $89.958 \pm 0.025$ \\
HAT-P-7B  & 2013 May 31 & \Ksfilt & $3859.3 \pm 1.7$ & $90.020 \pm 0.025$ \\
HAT-P-7B  & 2013 Jun 22 & \Ksfilt & $3858.7 \pm 2.0$ & $90.020 \pm 0.024$ \\
HAT-P-7B  & 2014 Jul 12 & \Ksfilt & $3858.6 \pm 1.7$ & $90.010 \pm 0.024$ \\
HAT-P-8B  & 2012 Jul 27 & \Kpfilt & $1037.9 \pm 1.5$ & $137.601 \pm 0.084$ \\
HAT-P-8C  & 2012 Jul 27 & \Kpfilt & $1047.8 \pm 1.6$ & $140.868 \pm 0.089$ \\
HAT-P-8B  & 2013 Aug 19 & \Kpfilt & $1040.3 \pm 1.6$ & $137.810 \pm 0.084$ \\
HAT-P-8C  & 2013 Aug 19 & \Kpfilt & $1045.0 \pm 1.5$ & $141.058 \pm 0.083$ \\
HAT-P-8B  & 2013 Aug 19 & \Ksfilt & $1041.0 \pm 1.5$ & $137.770 \pm 0.084$ \\
HAT-P-8C  & 2013 Aug 19 & \Ksfilt & $1044.7 \pm 1.5$ & $141.059 \pm 0.092$ \\
HAT-P-10B & 2012 Feb 02 & \Kpfilt & $342.5 \pm 1.5$ & $214.09 \pm 0.24$ \\
HAT-P-10B & 2013 Aug 19 & \Kpfilt & $355.0 \pm 1.5$ & $215.65 \pm 0.24$ \\
HAT-P-10B & 2013 Aug 19 & \Ksfilt & $355.0 \pm 1.5$ & $215.64 \pm 0.24$ \\
HAT-P-14B & 2012 Jun 05 & \Kpfilt & $857.6 \pm 1.5$ & $264.10 \pm 0.10$ \\
HAT-P-14B & 2013 Mar 26 & \Ksfilt & $857.4 \pm 1.5$ & $264.24 \pm 0.11$ \\
HAT-P-14B & 2014 Jul 07 & \Ksfilt & $856.9 \pm 1.6$ & $264.38 \pm 0.13$ \\
HAT-P-16B & 2012 Feb 02 & \Kpfilt & $691.6 \pm 1.5$ & $153.83 \pm 0.12$ \\
HAT-P-16B & 2013 Aug 19 & \Ksfilt & $688.6 \pm 1.5$ & $153.65 \pm 0.12$ \\
HAT-P-24B & 2012 Feb 02 & \Kpfilt & $4946.1 \pm 1.8$ & $170.894 \pm 0.020$ \\
HAT-P-24B & 2014 May 13 & \Ksfilt & $4944.2 \pm 1.9$ & $170.872 \pm 0.020$ \\
HAT-P-30B & 2012 Feb 02 & \Kpfilt & $3835.1 \pm 1.7$ & $4.219 \pm 0.024$ \\
HAT-P-30B & 2014 May 13 & \Ksfilt & $3836.6 \pm 1.7$ & $4.206 \pm 0.024$ \\
HAT-P-32B & 2012 Feb 02 & \Kpfilt & $2935.5 \pm 1.6$ & $110.583 \pm 0.030$ \\
HAT-P-32B & 2013 Mar 02 & \Ksfilt & $2935.5 \pm 1.7$ & $110.624 \pm 0.031$ \\
HAT-P-33B & 2012 Feb 02 & \Kpfilt & $307.2 \pm 1.5$ & $117.86 \pm 0.29$ \\
HAT-P-33B & 2013 Mar 02 & \Ksfilt & $306.3 \pm 1.5$ & $118.05 \pm 0.29$ \\
TrES-2B   & 2012 Jul 27 & \Kpfilt & $1105.4 \pm 1.5$ & $136.325 \pm 0.077$ \\
TrES-2B   & 2013 May 31 & \Ksfilt & $1106.7 \pm 1.5$ & $136.357 \pm 0.077$ \\
TrES-4B   & 2012 Jul 27 & \Kpfilt & $1562.3 \pm 1.6$ & $0.357 \pm 0.056$ \\
TrES-4B   & 2013 Jul 04 & \Ksfilt & $1563.4 \pm 1.5$ & $0.386 \pm 0.055$ \\
WASP-1B   & 2012 Jul 27 & \Kpfilt & $4580.0 \pm 1.8$ & $1.901 \pm 0.021$ \\
WASP-1B   & 2013 Aug 19 & \Ksfilt & $4581.5 \pm 1.8$ & $1.953 \pm 0.021$ \\
WASP-1B   & 2014 Jul 12 & \Ksfilt & $4582.3 \pm 1.8$ & $1.958 \pm 0.021$ \\
WASP-1B   & 2014 Oct 03 & \Ksfilt & $4581.7 \pm 1.8$ & $1.917 \pm 0.021$ \\
WASP-2B   & 2012 Jul 27 & \Kpfilt & $734.3 \pm 1.5$ & $104.39 \pm 0.12$ \\
WASP-2B   & 2013 Jun 22 & \Ksfilt & $730.0 \pm 1.5$ & $104.54 \pm 0.12$ \\
WASP-3B   & 2012 Jun 05 & \Kpfilt & $1192.2 \pm 1.6$ & $87.103 \pm 0.077$ \\
WASP-3B   & 2012 Jul 27 & \Kpfilt & $1191.0 \pm 1.5$ & $87.070 \pm 0.075$ \\
WASP-3B   & 2013 May 31 & \Ksfilt & $1189.1 \pm 1.6$ & $87.169 \pm 0.081$ \\
WASP-8B   & 2012 Jul 27 & \Kpfilt & $4505.2 \pm 1.7$ & $170.948 \pm 0.021$ \\
WASP-8B   & 2013 Aug 19 & \Ksfilt & $4507.4 \pm 2.0$ & $170.988 \pm 0.021$ \\
WASP-12B  & 2012 Feb 02 & \Kpfilt & $1058.8 \pm 1.5$ & $251.242 \pm 0.084$ \\
WASP-12C  & 2012 Feb 02 & \Kpfilt & $1067.1 \pm 1.5$ & $246.751 \pm 0.083$ \\
WASP-12B  & 2013 Mar 02 & \Ksfilt & $1058.6 \pm 1.5$ & $251.444 \pm 0.082$ \\
WASP-12C  & 2013 Mar 02 & \Ksfilt & $1068.4 \pm 1.5$ & $246.945 \pm 0.082$ \\
WASP-14B  & 2012 Jul 27 & \Kpfilt & $1450.8 \pm 1.5$ & $102.066 \pm 0.059$ \\
WASP-14B  & 2013 Mar 26 & \Kpfilt & $1449.1 \pm 1.5$ & $102.210 \pm 0.060$ \\
WASP-14B  & 2013 Mar 26 & \Ksfilt & $1449.5 \pm 1.5$ & $102.207 \pm 0.060$
\enddata
\tablecomments{We use both the \Kpfilt\ and \Ksfilt\ band for astrometric measurements. We imaged a few of our targets on the same night in both bands (HAT-P-8, HAT-P-10, and WASP-14) and show that the astrometry agrees to well within our uncertainties. The $\rho$ column shows the on-sky separation between the fitted positions of the primary and companion stars, in milliarcseconds. The PA column shows the position angle, in degrees east of north.}
\end{deluxetable}

\clearpage
\begin{deluxetable*}{llccl}
\tablecolumns{5}
\tablecaption{Astrometric measurements from previous studies for confirmed common proper motion companions \label{tab:prev_studies}}
\tablehead{
& & \colhead{$\rho$} & \colhead{PA} & \vspace{-0.1cm} \\
\colhead{Companion} & \colhead{UT Obs. Date} & \vspace{-0.2cm}  & & \colhead{Reference} \\
& & \colhead{(mas)} & \colhead{(\deg)} &
}
\startdata
HAT-P-7B & 2009 Jul 18-22\tablenotemark{a} & $3870 \pm   30$ & $ 90.4 \pm   0.5$ & \citet{Faedi2013}\\
HAT-P-7B & 2009 Aug 06 & $3875 \pm    5$ & $ 89.8 \pm   0.3$ & \citet{Narita2010b}\\
HAT-P-7B & 2009 Oct 30 & $3820 \pm   10$ & $ 90.4 \pm   0.1$ & \citet{Bergfors2013}\\
HAT-P-7B & 2009 Nov 02 & $3861 \pm    6$ & $ 89.8 \pm   0.1$ & \citet{Narita2012}\\
HAT-P-7B & 2011 Aug 12 & $3871 \pm    6$ & $ 89.7 \pm   0.1$ & \citet{Narita2012}\\
HAT-P-7B & 2011 Nov 09 & $3860 \pm   70$ & $ 89.9 \pm   0.8$ & \citet{Bergfors2013}\\
HAT-P-7B & 2012 Jul 07 & $3860 \pm    4$ & $ 89.8 \pm   0.1$ & \citet{Narita2012}\\
HAT-P-8B & 2012 Jun 24 & $1040 \pm   14$ & $137.9 \pm   0.8$ & \citet{Bechter2014}\\
HAT-P-8B & 2013 Jul 03 & $1053 \pm   14$ & $137.6 \pm   0.8$ & \citet{Bechter2014}\\
HAT-P-8C & 2012 Jun 24 & $1049 \pm   14$ & $141.4 \pm   0.8$ & \citet{Bechter2014}\\
HAT-P-8C & 2013 Jul 03 & $1041 \pm   14$ & $140.7 \pm   0.8$ & \citet{Bechter2014}\\
TrES-2B & 2007 May\tablenotemark{a} & $1089 \pm    8$ & $135.5 \pm   0.1$ & \citet{Daemgen2009}\\
TrES-2B & 2009 Jul 18-22\tablenotemark{a} & $1110 \pm   30$ & $137.0 \pm   2.0$ & \citet{Faedi2013}\\
TrES-2B & 2009 Oct 29 & $1085 \pm    6$ & $136.1 \pm   0.2$ & \citet{Bergfors2013}\\
TrES-4B & 2008 Jun\tablenotemark{a} & $1555 \pm    5$ & $ -0.2 \pm   0.1$ & Daemgen et al. (2009)\\
TrES-4B & 2009 Jul 18-22\tablenotemark{a} & $1540 \pm   30$ & $  1.2 \pm   1.2$ & \citet{Faedi2013}\\
TrES-4B & 2009 Oct 30 & $1550 \pm    7$ & $ -0.1 \pm   0.2$ & \citet{Bergfors2013}\\
WASP-2B & 2007 Nov\tablenotemark{a} & $ 757 \pm    1$ & $104.7 \pm   0.3$ & \citet{Daemgen2009}\\
WASP-2B & 2009 Apr 13 & $ 761 \pm    9$ & $103.5 \pm   0.2$ & \citet{Bergfors2013}\\
WASP-2B & 2009 Oct 29 & $ 739 \pm   24$ & $104 \pm   1.3$ & \citet{Bergfors2013}\\
WASP-2B & 2011 Nov 09 & $ 744 \pm   13$ & $104.6 \pm   0.7$ & \citet{Bergfors2013}\\
WASP-12BC\tablenotemark{b} & 2009 Oct 30 & $1047 \pm   21$ & $249.7 \pm   0.8$ & \citet{Bergfors2013}\\
WASP-12BC\tablenotemark{b} & 2011 Nov 8-9\tablenotemark{a} & $1043 \pm   14$ & $249.9 \pm   0.5$ & \citet{Bergfors2013}\\
WASP-12BC\tablenotemark{b} & 2011 Dec 04 & $1055 \pm   26$ & $250.0 \pm   1.0$ & \citet{Crossfield2012}\\
WASP-12BC\tablenotemark{b} & 2012 Feb 25 & $1078 \pm   33$ & $249.4 \pm   1.1$ & \citet{Crossfield2012}
\enddata
\tablenotetext{a}{A specific date was not reported. We report the date to the same precision as it appears in original work.}
\tablenotetext{b}{This study did not resolve the two companions. Thus, the measurement for the center of light of both companions is reported here and only used in the plots for the combined light of WASP-12BC.}
\tablecomments{Astrometric measurements of Friends of Hot Jupiters companions reported in previous studies. Unless otherwise stated, only studies reporting a numerical value with uncertainties for both separation and position angle of all companions are included in this table and plotted in Figure~\ref{fig:astr_plot}.}
\end{deluxetable*}

\begin{deluxetable*}{lccl}
\tablecolumns{4}
\tablecaption{Astrometric measurements from previous studies for candidate companions determined to be background objects\label{tab:astr_bg}}
\tablehead{
& \colhead{$\rho$} & \colhead{PA} & \vspace{-0.1cm} \\
\colhead{UT Obs. Date} & \vspace{-0.2cm}  & & \colhead{Reference} \\
& \colhead{(mas)} & \colhead{(\deg)} &
}
\startdata
\sidehead{\bf For the second candidate companion to HAT-P-7}
2009 Aug 06 & $3139 \pm 5$ & $ 266.30 \pm 0.37$ & \citet{Narita2010b}\\
2009 Nov 02 & $3137 \pm 6$ & $ 266.14 \pm 0.09$ & \citet{Narita2012}\\
2011 Aug 12 & $3103 \pm 11$ & $ 266.23 \pm 0.24$ & \citet{Narita2012}\\
2012 Jul 07 & $3095 \pm 4$ & $ 265.73 \pm 0.08$ & \citet{Narita2012}\\
2012 Jul 27 & $3091.5 \pm  3.3$ & $ 266.00 \pm 0.15$ & this work\\
2013 May 31 & $3083.0 \pm  2.7$ & $ 265.947 \pm 0.059$ & this work\\
2014 Jul 12 & $3065.5 \pm  1.7$ & $ 265.705 \pm 0.044$ & this work\\
\sidehead{\bf For the candidate companion to HAT-P-15}
2012 Feb 02 & $7099.3 \pm 9.9$ & $ 58.834 \pm 0.097$ & this work\\
2014 Oct 03 & $6382.6 \pm 7.70$ & $ 58.193 \pm 0.058$ & this work
\enddata
\tablecomments{These measurements are used to generate the plots in Figure~\ref{fig:astr_plot_bg}.}
\end{deluxetable*}

\clearpage
\begin{deluxetable*}{llccccccc}
\tablecolumns{9}
\tabletypesize{\scriptsize}
\tablecaption{Derived stellar parameters of confirmed common proper motion companions\label{tab:interp_sec}}
\tablehead{
& &  \colhead{$T_{\mathrm{eff}}$} & \colhead{$M$} & \colhead{$\log g$} & \colhead{$D$} & \colhead{\Jfilt-band $T_{\mathrm{eff}}$} & \colhead{\Hfilt-band $T_{\mathrm{eff}}$} & \colhead{\Kfilt-band $T_{\mathrm{eff}}$} \vspace{-0.1cm}\\
\colhead{Companion} & \colhead{UT Obs. Date} & \vspace{-0.2cm} & & & & & & \\
& & \colhead{(K)} & \colhead{(\Msun)} & \colhead{(cgs)} & \colhead{(AU)} & \colhead{(K)} & \colhead{(K)} & \colhead{(K)}
}
\startdata
HAT-P-7B & 2012 Jul 27 & $3321 \pm 28$ & $0.215 \pm 0.016$ & $5.047 \pm 0.018$ & $1238 \pm 67$ & $3291^{+14}_{-15}$ & $3357^{+16}_{-14}$ & $3316.7^{+11}_{-9.8}$ \\
HAT-P-7B & 2013 May 31 & $3307 \pm 10$ & $0.2089 \pm 0.0043$ & $5.0563 \pm 0.0047$ & $1238 \pm 67$ & \nodata & \nodata & $3307.0 \pm 6.5$ \\
HAT-P-7B & 2013 Jun 22 & $3320 \pm 30$ & $0.214 \pm 0.018$ & $5.047 \pm 0.019$ & $1238 \pm 67$ & \nodata & $3351^{+18}_{-17}$ & $3291^{+23}_{-24}$ \\
HAT-P-7B & 2014 Jul 12 & $3328 \pm 11$ & $0.2215 \pm 0.0066$ & $5.0428 \pm 0.0070$ & $1238 \pm 67$ & \nodata & \nodata & $3328.1^{+11}_{-9.8}$ \\
HAT-P-8B & 2012 Jul 27 & $3216 \pm 59$ & $0.168 \pm 0.020$ & $5.102 \pm 0.029$ & $236 \pm 13$ & $3153^{+31}_{-33}$ & $3293.2^{+9.8}_{-11}$ & $3208^{+16}_{-15}$ \\
HAT-P-8B & 2013 Aug 19 & $3252 \pm 41$ & $0.182 \pm 0.017$ & $5.086 \pm 0.022$ & $237 \pm 13$ & \nodata & $3306.2^{+5.7}_{-6.5}$ & $3216^{+20}_{-18}$ \\
HAT-P-8C & 2012 Jul 27 & $3058 \pm 63$ & $0.131 \pm 0.011$ & $5.167 \pm 0.023$ & $239 \pm 13$ & $2989^{+50}_{-42}$ & $3136^{+19}_{-18}$ & $3055^{+16}_{-18}$ \\
HAT-P-8C & 2013 Aug 19 & $3130 \pm 44$ & $0.146 \pm 0.010$ & $5.139 \pm 0.018$ & $238 \pm 13$ & \nodata & $3188^{+20}_{-18}$ & $3092^{+17}_{-15}$ \\
HAT-P-10B & 2012 Feb 02 & $3483 \pm 43$ & $0.344 \pm 0.047$ & $4.922 \pm 0.038$ & $41.7 \pm 1.4$ & $3525.4^{+3.9}_{-3.3}$ & \nodata & $3443.1^{+3.3}_{-2.6}$ \\
HAT-P-10B & 2013 Aug 19 & $3494 \pm 37$ & $0.362 \pm 0.035$ & $4.913 \pm 0.031$ & $43.2 \pm 1.5$ & \nodata & $3545.0^{+3.3}_{-2.6}$ & $3469.2^{+5.2}_{-3.9}$ \\
HAT-P-14B & 2012 Jun 05 & $3310 \pm 17$ & $0.211 \pm 0.010$ & $5.054 \pm 0.011$ & $175.8 \pm 9.4$ & \nodata & \nodata & $3310.2^{+5.7}_{-6.5}$ \\
HAT-P-14B & 2013 Mar 26 & $3356 \pm 57$ & $0.232 \pm 0.037$ & $5.021 \pm 0.039$ & $175.8 \pm 9.4$ & \nodata & $3413^{+15}_{-13}$ & $3302^{+18}_{-16}$ \\
HAT-P-14B & 2014 Jul 07 & $3263 \pm 20$ & $0.1873 \pm 0.0087$ & $5.082 \pm 0.011$ & $175.7 \pm 9.4$ & \nodata & \nodata & $3263.1^{+7.3}_{-6.5}$ \\
HAT-P-16B & 2012 Feb 02 & $3255 \pm 67$ & $0.182 \pm 0.027$ & $5.081 \pm 0.035$ & $162.5 \pm 6.9$ & $3319.2 \pm 7.3$ & \nodata & $3194.8^{+5.7}_{-4.9}$ \\
HAT-P-16B & 2013 Aug 19 & $3269 \pm 51$ & $0.188 \pm 0.023$ & $5.076 \pm 0.028$ & $161.8 \pm 6.9$ & \nodata & $3315.1^{+8.9}_{-8.1}$ & $3225 \pm 15$ \\
HAT-P-24B & 2012 Feb 02 & $3434 \pm 26$ & $0.298 \pm 0.026$ & $4.968 \pm 0.022$ & $1959 \pm 99$ & \nodata & \nodata & $3434^{+16}_{-19}$ \\
HAT-P-24B & 2014 May 13 & $3499 \pm 29$ & $0.373 \pm 0.030$ & $4.908 \pm 0.025$ & $1958 \pm 99$ & $3474.4^{+13}_{-9.8}$ & $3498.8^{+8.9}_{-8.1}$ & $3523.2 \pm 1.6$ \\
HAT-P-30B & 2012 Feb 02 & $3634 \pm 29$ & $0.489 \pm 0.020$ & $4.808 \pm 0.020$ & $740 \pm 31$ & \nodata & \nodata & $3634^{+13}_{-11}$ \\
HAT-P-30B & 2014 May 13 & $3692 \pm 40$ & $0.519 \pm 0.019$ & $4.777 \pm 0.019$ & $740 \pm 31$ & $3709.3 \pm 2.4$ & $3692.2 \pm 1.6$ & $3674^{+15}_{-13}$ \\
HAT-P-32B & 2012 Feb 02 & $3516 \pm 12$ & $0.393 \pm 0.012$ & $4.8930 \pm 0.0098$ & $831 \pm 15$ & $3524.0^{+5.7}_{-4.9}$ & \nodata & $3508.6 \pm 6.5$ \\
HAT-P-32B & 2013 Mar 02 & $3551 \pm 10$ & $0.4243 \pm 0.0085$ & $4.8677 \pm 0.0070$ & $831 \pm 15$ & \nodata & $3555^{+15}_{-14}$ & $3548^{+16}_{-15}$ \\
HAT-P-33B & 2012 Feb 02 & $3653 \pm 54$ & $0.493 \pm 0.031$ & $4.800 \pm 0.031$ & $118.9 \pm 2.8$ & $3705^{+20}_{-18}$ & \nodata & $3604^{+20}_{-18}$ \\
HAT-P-33B & 2013 Mar 02 & $3776 \pm 32$ & $0.557 \pm 0.014$ & $4.740 \pm 0.013$ & $118.5 \pm 2.8$ & \nodata & $3800^{+13}_{-11}$ & $3753^{+21}_{-20}$ \\
TrES-2B & 2012 Jul 27 & $3669 \pm 29$ & $0.509 \pm 0.013$ & $4.787 \pm 0.014$ & $216 \pm 13$ & $3651.6^{+7.3}_{-4.9}$ & $3701.2 \pm 2.4$ & $3655.7^{+6.5}_{-5.7}$ \\
TrES-2B & 2013 May 31 & $3662 \pm 39$ & $0.501 \pm 0.021$ & $4.793 \pm 0.021$ & $216 \pm 13$ & \nodata & $3697.9^{+2.4}_{-3.3}$ & $3627.2 \pm 8.1$ \\
TrES-4B & 2012 Jul 27 & $3985 \pm 93$ & $0.626 \pm 0.028$ & $4.677 \pm 0.023$ & $900 \pm 89$ & $4048.2 \pm 6.5$ & $4023.8 \pm 6.5$ & $3889^{+24}_{-21}$ \\
TrES-4B & 2013 Jul 04 & $3910 \pm 110$ & $0.602 \pm 0.036$ & $4.696 \pm 0.032$ & $900 \pm 89$ & \nodata & $4019.8^{+4.9}_{-3.3}$ & $3817 \pm 11$ \\
WASP-1B & 2012 Jul 27 & $3446 \pm 15$ & $0.313 \pm 0.019$ & $4.956 \pm 0.014$ & $1587^{+160}_{-16}$ & \nodata & \nodata & $3446.0^{+11}_{-8.9}$ \\
WASP-1B & 2013 Aug 19 & $3420 \pm 26$ & $0.285 \pm 0.023$ & $4.978 \pm 0.021$ & $1587^{+160}_{-16}$ & \nodata & $3441.9^{+6.5}_{-5.7}$ & $3398.0 \pm 8.1$ \\
WASP-1B & 2014 Jul 12 & $3395 \pm 17$ & $0.266 \pm 0.014$ & $4.998 \pm 0.013$ & $1587^{+160}_{-16}$ & \nodata & \nodata & $3394.8^{+11}_{-9.8}$ \\
WASP-1B & 2014 Oct 03 & $3442 \pm 18$ & $0.308 \pm 0.021$ & $4.960 \pm 0.017$ & $1587^{+160}_{-16}$ & \nodata & \nodata & $3442^{+13}_{-15}$ \\
WASP-2B & 2012 Jul 27 & $3509 \pm 26$ & $0.383 \pm 0.028$ & $4.899 \pm 0.022$ & $113.0 \pm 6.1$ & $3527.9 \pm 2.2$ & $3523.55^{+0.72}_{-1.4}$ & $3477.4^{+4.3}_{-3.6}$ \\
WASP-2B & 2013 Jun 22 & $3514 \pm 16$ & $0.391 \pm 0.016$ & $4.895 \pm 0.013$ & $112.3 \pm 6.1$ & \nodata & $3525.7^{+1.4}_{-2.2}$ & $3502.6 \pm 2.9$ \\
WASP-3B & 2012 Jun 05 & $2909 \pm 33$ & $0.1089 \pm 0.0035$ & $5.223 \pm 0.010$ & $300 \pm 22$ & \nodata & \nodata & $2909^{+17}_{-18}$ \\
WASP-3B & 2012 Jul 27 & $2871 \pm 51$ & $0.1053 \pm 0.0048$ & $5.233 \pm 0.015$ & $299 \pm 22$ & $2825 \pm 47$ & $2924^{+24}_{-28}$ & $2867^{+20}_{-18}$ \\
WASP-3B & 2013 May 31 & $2922 \pm 48$ & $0.1109 \pm 0.0058$ & $5.216 \pm 0.016$ & $299 \pm 22$ & \nodata & $2962^{+29}_{-25}$ & $2883.6 \pm 9.8$ \\
WASP-8B & 2012 Jul 27 & $3530 \pm 130$ & $0.34 \pm 0.11$ & $4.881 \pm 0.099$ & $383 \pm 48$ & $3383.3 \pm 9.6$ & $3568 \pm 12$ & $3666.0^{+6.4}_{-7.2}$ \\
WASP-8B & 2013 Aug 19 & $3657 \pm 76$ & $0.504 \pm 0.043$ & $4.793 \pm 0.043$ & $384 \pm 48$ & \nodata & \nodata & $3657.2^{+5.6}_{-6.4}$ \\
WASP-12B & 2012 Feb 02 & $3786 \pm 53$ & $0.561 \pm 0.022$ & $4.736 \pm 0.021$ & $462 \pm 39$ & $3810^{+36}_{-28}$ & \nodata & $3762^{+40}_{-35}$ \\
WASP-12B & 2013 Mar 02 & $3769 \pm 44$ & $0.554 \pm 0.020$ & $4.743 \pm 0.018$ & $462 \pm 39$ & \nodata & \nodata & $3768.6^{+11}_{-9.8}$ \\
WASP-12C & 2012 Feb 02 & $3748 \pm 90$ & $0.540 \pm 0.038$ & $4.753 \pm 0.037$ & $466 \pm 40$ & $3828.0 \pm 9.8$ & \nodata & $3674^{+12}_{-11}$ \\
WASP-12C & 2013 Mar 02 & $3808 \pm 49$ & $0.571 \pm 0.019$ & $4.727 \pm 0.017$ & $466 \pm 40$ & \nodata & \nodata & $3807.6^{+7.3}_{-6.5}$ \\
WASP-14B & 2012 Jul 27 & $3464 \pm 31$ & $0.331 \pm 0.037$ & $4.939 \pm 0.028$ & $302 \pm 23$ & $3476.0 \pm 3.3$ & $3484.2^{+4.1}_{-3.3}$ & $3432.2^{+4.9}_{-3.3}$ \\
WASP-14B & 2013 Mar 26 & $3464 \pm 19$ & $0.336 \pm 0.024$ & $4.939 \pm 0.018$ & $302 \pm 23$ & \nodata & \nodata & $3461.4^{+8.1}_{-7.3}$ 
\enddata
\tablecomments{The $T_{\mathrm{eff}}$, $M$, $\log g$, and $D$ columns are error weighted averages from all measurements of each target for each date. These uncertainties include measurement uncertainty, uncertainties from the primary star's stellar parameter and uncertainties on the error weighted average but do not include uncertainties arising from the stellar models and our assumptions on stellar composition. The uncertainties in last three columns, which report $T_{\mathrm{eff}}$ for each filter, only include measurement uncertainties and thus would be underestimates of the true uncertainty. The last three columns show that the companion object have $T_{\mathrm{eff}}$ measurements consistent with a late type main sequence star in all filters.}
\end{deluxetable*}

\clearpage
\begin{deluxetable*}{lcccl}
\tablecolumns{5}
\tablecaption{Stellar companion fractions from other surveys \label{tab:other_surveys}}
\tablehead{
\colhead{Survey} & \colhead{$\eta$} & \colhead{Range} & \colhead{$N_{\mathrm{targets}}$} & \colhead{Notes}
}
\startdata
\sidehead{\bf This work}
Total sample, uncorrected & $34\% \pm 7\%$ & 5.5\arcsec & 50 & Diffraction-limited AO\\
Total sample, corrected & $49\% \pm 9\%$ & 5.5\arcsec & 50 & Diffraction-limited AO\\
Misaligned sample, corrected & $48\% \pm 12\%$ & 5.5\arcsec & 27 & Diffraction-limited AO\\
Control sample, corrected & $51\% \pm 13\%$ & 5.5\arcsec & 23 & Diffraction-limited AO\\
\sidehead{\bf Near-infrared direct imaging surveys of transiting gas giant planet hosts}
\citet{Daemgen2009} & $21\% \pm 12\%$ & Unknown\tablenotemark{a} & 14 & Lucky imaging\\
\citet{Faedi2013} & $38\% \pm 15\%$ & Unknown\tablenotemark{a} & 16 & Lucky imaging\\
\citet{Bergfors2013} & $29\% \pm 12\%$ & Unknown\tablenotemark{a} & 21 & Lucky imaging\\
\citet{Adams2013} & $33\% \pm 15\%$ & Unknown\tablenotemark{a} & 15 & Diffraction-limited AO\\
\sidehead{\bf Near-infrared direct imaging surveys of {\it Kepler} planet candidate hosts}
\citet{Adams2012} & $59\% \pm 5\%$ & 6\arcsec\ & 90 & Diffraction-limited AO\\
\citet{Adams2013} & $33\% \pm 17\%$ & Unknown\tablenotemark{a} & 12 & Diffraction-limited AO\\
\citet{Dressing2014} & $31\% \pm 5\%$ & 4\arcsec\ & 87 & Diffraction-limited AO\\ 
\citet{LilloBox2012,LilloBox2014} & $33\% \pm 4\%$ & 6\arcsec\ & 174 & Lucky imaging\\ 
\citet{Wang2014} & $45\% \pm 7\%$ & 6\arcsec\ & 56 & Diffraction-limited AO\\
\sidehead{\bf Optical direct imaging surveys of {\it Kepler} planet candidate hosts}
\citet{Law2014} & $7.4\% \pm 1.0\% $ & 2.5\arcsec\ & 715 & Diffraction-limited AO\\
 & $14.0^{+5.8}_{-3.5}\%$ & & & Only hosts of planets $P<15$ days, $R_p > R_{\mathrm{Neptune}}$\\
 \citet{Gilliland2015} &  $22\% \pm 10\%$ & 2.5\arcsec\ & 23 & Hubble Space Telescope WFC3 imaging\\
 \citet{Horch2014}, raw & $7.0\% \pm 1.1\%$ & 1\arcsec\ & 588 & Differential speckle survey on WIYN 3.5m telescope\\
 & $23\% \pm 8\%$ & 1\arcsec\ & 35 & Differential speckle survey on Gemini-N 8.1m telescope\\
  \citet{Horch2014}, corrected & $37\% \pm 7\%$ & 1\arcsec\ & 588 & Differential speckle survey on WIYN 3.5m telescope\\
 & $47\% \pm 19\%$ & 1\arcsec\ & 35 & Differential speckle survey on Gemini-N 8.1m telescope\\
\sidehead{\bf Multi-modal surveys of field stars}
\citet{Raghavan2010} & $44\% \pm 2\%$ & varies\tablenotemark{b} & 454 & Solar-like stars with 25\unit{pc}\\
  & $24\% \pm 1\%$ & See note\tablenotemark{c} & 454 & For comparison with our survey\\
\enddata
\tablenotetext{a}{These studies do not report a number of companions in any given range.}
\tablenotetext{b}{This study uses multiple techniques to find stellar companions, each with different sensitivities for companion separation.}
\tablenotetext{c}{This line reports the fraction of stars with companions in the same range as our survey, as detailed in Section~\ref{sec:compare_field}.}
\tablecomments{Unless otherwise noted, companion fraction $\eta$ reported are based on raw counts, with no correction for completeness. The uncertainties for completeness corrected companion fractions do not scale inversely with the square root of the number of targets because the completeness correction effectively multiplies both the value and uncertainty by a constant.}
\end{deluxetable*}

\end{document}